\documentclass[a4paper,11pt]{article}
\pdfoutput=1

\usepackage{jheppub}

\usepackage[utf8]{inputenc}
\usepackage[dvipsnames,table,xcdraw]{xcolor}
\usepackage{graphicx}
\usepackage{amsmath}
\usepackage{amssymb,amsthm,physics}
\usepackage{slashed}
\usepackage{listings}
\usepackage{color}
\usepackage{orcidlink}
\usepackage{listings}
\usepackage{caption}
\usepackage{subcaption}
\usepackage{fancyheadings}
\usepackage[title,titletoc]{appendix}

\numberwithin{equation}{section}

\def\nn{\nonumber \\}

\def\I{\mathcal{I}}

\def\O{\mathcal{O}}

\def\z{\mathbf{z}}
\def\x{\mathbf{x}}

\def\g{\mathbf{g}}
\def\f{\mathbf{f}}

\def\F{\mathcal{F}}
\def\U{\mathcal{U}}

\definecolor{shaded}{RGB}{252,252,252}
\preprint{{\raggedleft%
ZU-TH 44/25
}}

\title{\textsc{CALICO}: Computing Annihilators from Linear Identities Constraining (differential) Operators}
\author[a,\,\orcidlink{0009-0008-3070-1250}]{Giuseppe Bertolini}
\author[b,\,\orcidlink{0000-0002-7543-7376}]{Gaia Fontana}
\author[a,\, \orcidlink{0000-0003-1534-4378}]{Tiziano Peraro}
\date{April 2024}

\affiliation[a]{Dipartimento di Fisica e Astronomia, Universit\'{a} di Bologna e INFN, Sezione di Bologna, via
  Irnerio 46, I-40126 Bologna, Italy}
\affiliation[b]{Physik-Institut, Universit\"{a}t Z\"{u}rich, Winterthurerstrasse 190, CH-8057 Z\"{u}rich, Switzerland}
\emailAdd{giuseppe.bertolini@studio.unibo.it}
\emailAdd{gaia.fontana@physik.uzh.ch}
\emailAdd{tiziano.peraro@unibo.it}

\abstract{
  We elaborate on the method of parametric annihilators for deriving integral relations. Parametric annihilators are differential operators that annihilate multivalued integration kernels appearing in suitable integral representations of special functions. We illustrate this approach in a way that applies to a broad variety of integral representations. We describe a method for computing parametric annihilators based on efficient linear solvers and use them to derive relations between a wide class of special functions related to important problems in high-energy physics.
  We also formulate a similar method for deriving differential equations satisfied by the independent integrals within an integral family.
    We show applications to several classes of special functions, including hypergeometric functions, loop integrals in various representations (including Baikov, loop-by-loop Baikov, Lee-Pomeransky and Schwinger representations) and duals of loop integrals.
We finally present the public \textsc{Mathematica} package \textsc{CALICO} for computing parametric annihilators and its usage in several examples of high relevance in theoretical particle physics.
}

\begin{document}

\maketitle

\section{Introduction}

Loop integrals are crucial ingredients for our theoretical understanding of quantum field theory. They contribute to precise phenomenological predictions about the fundamental interactions of nature, which are essential to match the uncertainty of modern experiments. They are also used for new approaches to classical theories --- including gravity --- which are increasingly important due to recent improvements in experimental observations. Due to the rich mathematical structure they exhibit, they are also an interesting subject for mathematical studies, which in turn we may exploit to improve our capability of making higher-order theoretical predictions in perturbation theory.

The set of loop integrals which contribute to a theoretical prediction are generally not independent. They satisfy linear relations, such as symmetry relations and, most importantly, \emph{Integration By Parts}~(\emph{IBP}) identities~\cite{Tkachov:1981wb,Chetyrkin:1981qh}, which can be found via purely algebraic means. By finding and solving such relations, this set of integrals is reduced to a linear combination of an independent subset of them, called \emph{master integrals}~(\emph{MIs}). The reduction to master integrals is an essential ingredient of most higher-loop theoretical predictions. Besides drastically reducing the number of integrals which need to be computed, it casts amplitudes in a form where many of its properties and symmetries are manifest, and it enables the method of differential equations~\cite{Kotikov:1990kg,Gehrmann:1999as} for computing the master integrals themselves.

IBPs are traditionally found using the momentum-space representation of loop integrals. However, many other representations of these integrals are known, including the Feynman, Schwinger, Baikov~\cite{Baikov:1996iu} (and its loop-by-loop variant~\cite{Frellesvig:2017aai}) and Lee-Pomeransky~\cite{Lee:2013hzt} representations. Different representations offer different tradeoffs, hence their usefulness is highly dependent on the problem at hand.

The Baikov representation, for instance, is used --- among other things --- in modern approaches to IBP reduction, combined with syzygy techniques, for finding simpler identities among loop integrals which do not contain higher-powers of denominators~\cite{Larsen:2015ped}. The Lee-Pomeransky representation has also been used in the context of integral reduction, most notably using the approach of \emph{parametric annihilators}~\cite{Baikov:1996iu,Lee:2014tja,Bitoun:2017nre}. The latter proved to be extremely efficient in certain contexts, e.g.\ for reducing high powers of denominators (see e.g.~\cite{vonManteuffel:2019wbj}) which often appear in integrals having desirable properties, such as quasi-finiteness~\cite{Panzer:2014gra,vonManteuffel:2014qoa} or a pure functional form~\cite{Henn:2013pwa}.  More recently, integral reduction in the Feynman representation has also be considered~\cite{Chen:2019mqc}.

The method of parametric annihilator has been extensively described in~\cite{Bitoun:2017nre} for \emph{twisted Mellin integrals} and in particular for the Lee-Pomeransky parametrization of loop integrals. It consists in finding \emph{differential operators} in the integration variables which annihilate the multivalued integration kernel appearing in such integral representation. These operators are thus used to find linear identities obeyed by loop integrals.

The main goals of this work are the following. We review and elaborate on the method of \emph{parametric annihilators} for finding integral identities, with a focus on parametric representations of loop integrals. We formulate the method in more general terms, extending its range of applications to different integral representations --- including Schwinger and loop-by-loop Baikov which had not been used in this context. In particular, we find the Schwinger representation yields very efficient systems of identities in some cases. We also illustrate a similar technique for finding differential equations satisfied by the master integrals. We provide a description of these methods in a language that is accessible to the scientific community of theoretical particle physics and phenomenology, in particular the one focusing on higher-order perturbative corrections, avoiding mathematical language that is not common knowledge in these fields. We describe and provide an implementation of annihilators and other important differential operators based on modern linear solvers which rely on cutting edge finite-field techniques~\cite{vonManteuffel:2014ixa,Peraro:2016wsq}. This implementation is distributed as a \textsc{Mathematica} package named \textsc{CALICO}, which we believe can be a useful tool for studying a broad variety of integrals, including loop integrals in various parametric representations. Finally, we introduce a new way of characterizing and finding relations between duals of loop integrals~\cite{Caron-Huot:2021xqj,Caron-Huot:2021iev,Fontana:2023amt} --- which play a crucial role in the study of loops within the framework of intersection theory~\cite{Mizera:2017rqa,Mastrolia:2018uzb} --- compatible with the approach discussed in this paper.

The paper is structured as follows. In section~\ref{sec:famsann}, we set some notation and describe the method of parametric annihilators in a way that covers a wide class of functions having a suitable integral representation. We also clarify the connection between this approach and others based on syzygy equations. In section~\ref{sec:de}, we describe two methods for deriving differential equations, the second of which also relies on finding suitable differential operators in the integration variables. In section~\ref{sec:comp}, we illustrate a method for computing annihilators by solving linear constraints, which is suitable for applying modern linear solvers and finite-field methods. In sections~\ref{sec:loop} to~\ref{sec:twisted}, we describe several applications of this approach, including different representations of loop integrals --- namely the standard and loop-by-loop Baikov, Lee-Pomeransky and Schwinger representations --- as well as duals of loop integrals. In section~\ref{sec:calico}, we describe the basic usage of the \textsc{CALICO} package. Finally, in section~\ref{sec:conclusions} we draw our conclusions and in appendix~\ref{sec:feynLPpar} we provide more details on the definitions of the polynomials which appear in the integral representations we use.

\section{Integral families and parametric annihilators}
\label{sec:famsann}

In this section we set up our notation and review the main concepts we use in this paper, namely parametric annihilators and their usage in finding linear identities within integral families.

\subsection{Basic definitions and notation}
\label{sec:def}

We use a multi-index notation.  Given an $n$-dimensional multi-index $\alpha=(\alpha_1,\ldots,\alpha_n)$ and a list of $n$ variables $\z=(z_1,\ldots,z_n)$, we define the monomial
\begin{equation}
  \z^\alpha = \prod_{j=1}^n z_j^{\alpha_j}
\end{equation}
and its total degree
\begin{equation}
  |\alpha| = \sum_{j=1}^n \alpha_j.
\end{equation}
When this does not cause ambiguity, we define the derivative operator as
\begin{equation}
  \partial_j \equiv \frac{\partial}{\partial z_j}.
\end{equation}

We consider an \emph{integral family}, whose elements are linear combinations of integrals of the type
\begin{equation}
    I_{\alpha} = \int \dd^n \z\,  \varphi_{\alpha}(\z)\, u(\z), \label{eq:Ialpha}
\end{equation}
where $u(\z)$ is a \emph{multivalued function} common to all integrals, also known as \emph{twist}.  In most of our applications, $u$ takes the form
\begin{equation}
  u(\z) = \prod_j B_j(\z)^{\gamma_j}, \label{eq:u}
\end{equation}
where $B_j(\z)$ are polynomials and $\gamma_j$ are \emph{generic exponents}, which typically depend on regulators.  Other forms are however possible, such as
\begin{equation}
  u(\z) = \exp F(\z)\, \prod_j B_j(\z)^{\gamma_j}, \label{eq:uexp}
\end{equation}
where $F(z)$ is a polynomial or a rational function.

The functions $\varphi_{\alpha}(\z)$ in eq.~\eqref{eq:Ialpha} are identified by the multi-index $\alpha$ and contain $n$ \emph{rational factors} raised to the \emph{integer exponents} $\alpha = (\alpha_1, \dots, \alpha_n)$.  In most cases $\varphi_{\alpha}$ is a rational function.
We assume the family of functions $\varphi_{\alpha}$ to be closed under differentiation and multiplication by monomials.  In other words, both $\partial_{j}\varphi_{\alpha}(\z)$ and $\z^\beta \varphi_{\alpha}(\z)$ are linear combinations of functions of the same form as $\varphi_{\alpha}(\z)$.  Explicit examples are given in later sections.

The integration domain never varies in our applications and is thus left implicit in this section.  We also understand that $u$ and $\varphi_\alpha$ may also depend on additional free parameters (other than regulators).  In the context of loop integrals, for instance, these are generally kinematic variables.  If $x$ is one of these free parameters, we assume that $\partial_{x}\varphi_{\alpha}(\z)$ either vanishes or belongs to the same space of functions $\varphi_{\alpha}$.  The integrals $I_\alpha$ are thus \emph{special functions} of the regulators and these parameters.

In this paper, we are interested in finding and solving \emph{linear relations} satisfied by integrals having the general form in eq.~\eqref{eq:Ialpha}.
An important ingredient, in this context, are \emph{Integration by parts identities} (\emph{IBPs}), which read
\begin{equation}
  \int \dd^n \z\,  \left( \partial_j \varphi_{\alpha}(\z)  \right)\, u(\z) + \int \dd^n \z\,  \varphi_{\alpha}(\z)\, \left( \partial_j \log u(\z)  \right)\, u(\z) = \int \dd^n \z\,   \partial_j \Big(  \varphi_{\alpha}(\z)\, u(\z)  \Big).
\end{equation}
The right-hand side (r.h.s.)\ is a boundary term.  Even in cases where the boundary term does not vanish, it can be rewritten as an integral in fewer integration variables and thus belonging to a simpler integral family.  In all our applications, the boundary term can be set to zero, hence it will be neglected in the remainder of this paper.  By carrying out derivatives explicitly on the left-hand side (l.h.s.)\ of the previous equation, we obtain non-trivial integral identities.  However, in most applications, the term $\partial_j \log u(\z)$ contains denominator factors which cannot be absorbed in the family of functions $\varphi_\alpha(\z)$, hence the l.h.s.\ is generally not a relation within the original integral family.  When $u(\z)$ has the form in~\eqref{eq:u}, as an example, the term $\partial_j \log u(\z)$ generates rational functions with $B_j(\z)$ as denominator factor or, equivalently, integrals with shifted exponents $\gamma_j\to \gamma_j-1$, also known as \emph{shifted integrals} (or dimensionally shifted integrals in the context of loop integrals).  In principle, we can deal with these by writing down \emph{recurrence relations}~\cite{Derkachov:1990osa,Tarasov:1996br,Lee:2010wea}, which relate integrals with shifted exponents.  It is generally more convenient, however, to write down identities for integrals within the same family and without shifted exponents.  In the following, we review how to achieve this by using the method of parametric annihilators.  We will expand on recurrence relations again in section~\ref{sec:de}.

\subsection{Integral identities via annihilators}
\label{sec:ann}
In this section we review the method of \emph{parametric annihilators} (henceforth simply annihilators, for brevity).  These are operators that annihilate integration kernels of special functions, including various parametric representations of loop integrals, which can be used to derive linear relations between them (see e.g.~\cite{Baikov:1996iu,Lee:2014tja,Bitoun:2017nre}).

An \emph{annihilator} of order $o$ of $u(\z)$ is an operator of the form
\begin{equation}
  \hat A = c_0(\z) + \sum_j c_j (\z)\, \partial_j + \sum_{j_1\leq j_2}c_{j_1 j_2}(\z)\, \partial_{j_1}\partial_{j_2}+\cdots+\sum_{j_1\leq\cdots\leq j_o}c_{j_1\cdots j_o}(\z)\, \partial_{j_1}\cdots\partial_{j_o} \label{eq:Ao}
\end{equation}
with \emph{polynomials} $c_{j_1 j_2\cdots}$ in $\z$, such that
\begin{equation}
  \hat A\, u(\z) = 0. \label{eq:A}
\end{equation}
For any annihilator $\hat A$, we have infinitely many integral
identities
\begin{equation}
  \int \dd^n \z\,  \varphi_{\alpha}(\z)\, \hat A \, u(\z) = 0
\end{equation}
for any multi-index $\alpha$.  We thus obtain non-trivial identities within the integral family in eq.~\eqref{eq:Ialpha} by integrating by parts derivatives.  More explicitly, neglecting boundary terms, we obtain
\begin{equation}
  \int u\, (\varphi_{\alpha} c_0) - \sum_j \int u\, (\partial_j c_j \varphi_\alpha) +\cdots+(-1)^o\sum_{j_1\leq\cdots\leq j_o}\int u\, \partial_{j_1}\cdots\partial_{j_o}(c_{j_1\cdots j_o} \varphi_\alpha) = 0, \label{eq:Atmpeq}
\end{equation}
where we omitted the integration measure and the $\z$ dependence for brevity.  Under our assumptions, all integrals in eq.~\eqref{eq:Atmpeq} belong to the integral family in~\eqref{eq:Ialpha}.

In general, for each annihilator $\hat A$, we use eq.~\eqref{eq:Atmpeq} to write a \emph{template identity}, which is an identity valid for generic, symbolic exponents~$\alpha$.  At a later stage, identities valid for specific integrals are obtained from the template identities, by replacing the symbolic $\alpha$ with a list of integer exponents.  A multi-index of integer exponents $\alpha$, or the function $\varphi_\alpha$ and integral $I_\alpha$ it identifies, is called \emph{seed} or \emph{seed integral} in this context.  A common strategy consists in applying each template identity to a large number of seed integrals and thus obtain a linear system of equations.\footnote{In the context of loop integrals, this is known as the Laporta algorithm~\cite{Laporta:2000dsw}.}  Additional relations, such as \emph{symmetry} relations, may also exist.  These are changes of integration variables which map integrals into linear combinations of integrals of the same family.  In applications where more than one integral family is considered, symmetry relations relating integrals of different families may also exist.

By solving this system of linear relations, one can reduce all integrals within a family (or several families) to a linear combination of a smaller subset of independent integrals, called \emph{master integrals} (\emph{MIs})
\begin{equation}
  I_\alpha = \sum_{\beta\in\textrm{MIs}} c_{\alpha\beta} I_\beta,\label{eq:red}
\end{equation}
where, with a slight abuse of notation in the summation, we confuse the (master) integrals with the multi-index that identifies them.
The reduction to MIs is a crucial step in many theoretical and phenomenological studies in particle physics and other fields, including classical gravity and cosmology.

It should be noted that, whenever a system is solved, an \emph{order} must be introduced between the unknowns, which determines which unknowns are substituted with higher priority. In this context, such an order is expressed in terms of the so-called \emph{weight}, which can be regarded as an estimate of the complexity of the integrals~\cite{Laporta:2000dsw}. While solving the system, more complex integrals (i.e.\ with higher weight) are eliminated in favour of simpler integrals (i.e.\ with lower weight). The list of MIs also depends on this weight, although their number does not. For a choice of weight used in several modern calculations see e.g.~\cite{Peraro:2019svx}.

We now review a few basic properties of annihilators.  Linear combinations of annihilators are obviously also annihilators.  Moreover, if $\hat A$ is an annihilator of $u(\z)$ then
\begin{equation}
  \z^\beta \, \hat A
\qquad \textrm{and} \qquad
  \partial_j \hat A \label{eq:zbdA}
\end{equation}
are also annihilators.  Using the language of algebraic geometry, this means than the set of annihilators of $u(\z)$ is a $D$-module.  In the second expression, we understand that the derivative $\partial_j$ acts on everything on its right, namely both the polynomial coefficients inside $\hat A$ and the function $\hat A$ is applied to.  However, we can easily see that these two annihilators are not needed, since
\begin{align}
  \int \dd^n \z\,  \varphi_{\alpha}(\z)\, \Big( \z^\beta \hat A \, u(\z)  \Big) ={}& \int \dd^n \z\,  \Big(\z^\beta \varphi_{\alpha}(\z)   \Big)\, \hat A \, u(\z)\nn
                                                                                         \int \dd^n \z\,  \varphi_{\alpha}(\z)\, \Big( \partial_j \hat A \, u(\z)  \Big) ={}& -\int \dd^n \z\,  \Big(\partial_j \varphi_{\alpha}(\z)   \Big)\, \hat A \, u(\z).
\end{align}
Since, under our assumptions, the space of functions $\varphi_{\alpha}(\z)$ is closed under monomial multiplication and differentiation, all the identities we can obtain from $\z^\beta \hat A$ and $\partial_j \hat A$ with seed $\varphi_{\alpha}(\z)$ can also be obtained as identities generated by $\hat A$ applied to a suitable set of seeds which span $\z^\beta \varphi_{\alpha}(\z)$ and $\partial_j \varphi_{\alpha}(\z)$.  Hence, we are generally interested in finding a set of \emph{generators} that is minimal, namely a set of annihilators $\hat A$ that are independent modulo linear combinations of operators obtained as in eq.~\eqref{eq:zbdA}.

\subsection{Connection with syzygy-based approaches}
\label{sec:compsyz}

Before we further elaborate on the method of annihilators and related techniques, we make an observation about the relation between this and another method that is commonly used in the literature to avoid the appearance of shifted integrals. The latter relies on syzygy equations.

For simplicity, we focus on a twist of the form
\begin{equation}
  u(\z) = B(\z)^\gamma, \label{eq:baikovliketwist}
\end{equation}
i.e.\ defined by just one polynomial $B(\z)$.  The most general IBP relation one can write for an integral of the corresponding family reads
\begin{align}
  0 ={}& \sum_{j=1}^n \int \dd^n\z\, \partial_j \Big( a_j(\z) \varphi_\alpha(\z) B(\z)^\gamma \Big) \nn
  ={}& \int \dd^n\z\, \left[  B(\z)^\gamma\, \sum_{j=1}^n\ \partial_j \Big( \varphi_\alpha(\z) a_j(\z) \Big) +  B(\z)^\gamma \varphi_\alpha(\z) \Big(\gamma\,  \frac{1}{B(\z)} \sum_{j=1}^n a_j(\z) \partial_j B(\z)  \Big)  \right], \label{eq:baikovibp}
\end{align}
where $a_j(\z)$ are polynomials in the integration variables.  The first term in the second equality, under our assumptions, is a linear combination of integrals within our family, since the space of functions $\varphi_\alpha$ is closed under differentiation and monomial multiplication.  The second term, instead, is proportional to a shifted integral, with $\gamma\to \gamma-1$, due to the denominator $1/B(\z)$.  The denominator however simplifies if a special choice of $a_j(\z)$ is made, namely such that
\begin{equation}
  \sum_{j=1}^n a_j(\z) \partial_j B(\z) = b(\z)\, B(\z)\label{eq:baikovnoshift}
\end{equation}
for some polynomial $b(\z)$. This is a \emph{syzygy equation} for the unknown polynomials $a_j(\z)$ and $b(\z)$, which can be solved with methods of algebraic geometry (we will further elaborate on this in section~\ref{sec:comp}).  After plugging a solution of eq.~\eqref{eq:baikovnoshift} into eq.~\eqref{eq:baikovibp}, we obtain and identity that is free of shifted integrals
\begin{equation}
  \int \dd^n\z\, B(\z)^\gamma\, \left[ \sum_{j=1}^n\ \partial_j \Big( \varphi_\alpha(\z) a_j(\z) \Big) +  \gamma \varphi_\alpha(\z) b(\z)  \right] = 0.\label{eq:baikovibpnoshift}
\end{equation}
This method has been used with the Baikov representation~\cite{Larsen:2015ped} and the Lee-Pomeranski representation~\cite{Sameshima:2019qpr} of loop integrals, both of which have the form in eq.~\eqref{eq:baikovliketwist}, although one can easily generalize it to a more complex twist.

The strategy we just reviewed is equivalent to the method of parametric annihilators when the latter is restricted to first-order operators~\cite{Bertolini}.  First, consider a first-order operator ($o=1$) of the form in eq.~\eqref{eq:Ao}.  Multiplying eq.~\eqref{eq:A} by $B(\z)^{1-\gamma}$, for our choice of twist, we obtain that the condition the annihilator must satisfy is equivalent to the following relation for its polynomial coefficients $c_j(\z)$
\begin{equation}
  c_0(\z)\, B(\z) + \gamma \sum_{j=1}^n c_j(\z)\, \partial_j B(\z) = 0.
\end{equation}
By comparing this relation with eq.~\eqref{eq:baikovnoshift} we immediately see that there is a one-to-one correspondence between first-order annihilators and syzygy solutions of~\eqref{eq:baikovnoshift}, via the identifications
\begin{equation}\label{eq:syzto1ann}
  b(\z) = -c_0(\z), \qquad a_j(\z) = \gamma\, c_j(\z).
\end{equation}
Second, by inserting this relation into eq.~\eqref{eq:baikovibpnoshift}, we obtain the same first-order integral identity we get from eq.~\eqref{eq:Atmpeq} by setting $o=1$. This shows that the syzygy method yields the same integral identities as first-order annihilators.

Additional syzgy equations can be used to further constrain the form of the identities that are generated, e.g.\ such that certain kinds of integrals do not appear~(see e.g.~\cite{Gluza:2010ws,Ita:2015tya,Larsen:2015ped,Wu:2023upw}). These additional constraints can be applied to both the solutions of eq.~\eqref{eq:baikovnoshift} and the ones of eq.~\eqref{eq:A} regardless of the order $o$ of the annihilator.

Hence, the method of parametric annihilators \emph{generalizes} the one based on syzgy equations. In cases where first-order annihilators are sufficient for the solution of a problem, the two strategies are equivalent. This seems to be the case when the standard Baikov representation is used, as we are not aware of counterexamples. When second-order or higher-order annihilators are needed (examples in the Lee-Pomeransky representation have already been reported~\cite{vonManteuffel:2019wbj} and more, for various representations, will be given in this work), the method of annihilators superseeds the one based on syzgy equations.  The two methods may become equivalent, if the syzygy method is generalized to include higher-order derivatives in eq.~\eqref{eq:baikovibp}, but this goes beyond the way it is commonly formulated.

\section{Differential equations}
\label{sec:de}

As we already mentioned, integrals in the form of eq.~\eqref{eq:Ialpha} typically also depend on additional free parameters.  A very effective approach for both studying the analytic structure of these integrals and evaluating them either analytically or numerically is the method of \emph{differential equations} (\emph{DEs})~\cite{Kotikov:1990kg,Gehrmann:1999as}.

Let $x$ be a free parameter the integrals depend on.  By reducing the derivative of MIs with respect to $x$ to MIs, we can write a system of differential equations satisfied by the MIs themselves
\begin{equation}
  \label{eq:de}
  \partial_x I_\alpha = \sum_{\beta\in\textrm{MIs}}M_{\alpha \beta} I_\beta, \qquad \textrm{for } \alpha\in \textrm{MIs}.
\end{equation}
In the following, we describe two strategies for deriving DEs.  The first is based on \emph{recurrence relations}, while the second is based on building suitable \emph{differential operators} in the integration variables.

\subsection{Differential equations via recurrence relations}

The most straightforward way of computing DEs for integrals of the form in eq.~\eqref{eq:Ialpha} is by differentiating directly under the integral sign, which leads to the need of recurrence relations.  For simplicity, we consider a twist $u$ of the form of eq.~\eqref{eq:baikovliketwist},
but generalizations to other forms of $u$ are straightforward.  By
differentiating at the integrand level we obtain
\begin{equation}
  \partial_x I_\alpha = \int \dd^n \z\,  \Big(\partial_x \varphi_{\alpha}(\z)\Big)\, B(\z)^\gamma + \gamma\, \int \dd^n \z\,  \varphi_{\alpha}(\z)\, \Big( \partial_x B(\z)  \Big)\, B(\z)^{\gamma-1}.\label{eq:dIalpharec}
\end{equation}
Under the assumptions outlined in section~\ref{sec:def}, the first term of the r.h.s.\ is an integral belonging to the original integral family.  The second term, however, belongs to an integral family of \emph{shifted} integrals, i.e.\ a family obtained from the original one by shifting exponents, namely $\gamma\to \gamma-1$.  In order to reduce the r.h.s.\ to a linear combination of integrals in the original family, with unshifted exponents, we need to derive a recurrence relation that shifts back the exponents from $\gamma-1$ to $\gamma$.  While in some cases there are closed formulas for such recurrence relation~\cite{Derkachov:1990osa,Tarasov:1996br,Lee:2010wea}, in the most general case a direct relation of this form is not available.  Deriving a recurrence relation for the opposite shift, namely from $\gamma+1$ to $\gamma$, is however straightforward.  Indeed, it simply consists in multiplying the function $\varphi_\alpha$ by the polynomial $B$:
\begin{equation}
  I_\alpha^{(\gamma+1)} = \int \dd^n \z\,  \Big(B(\z) \varphi_{\alpha}(\z)\Big)\, B(\z)^\gamma,
\end{equation}
where $I_\alpha^{(\gamma')} \equiv I_\alpha|_{\gamma\to\gamma'}$.  By writing such a relation for all MIs and reducing the r.h.s.\ of it back to MIs we obtain a matrix $R^{(+)}$ which realizes the recurrence
\begin{equation}
  I_\alpha^{(\gamma+1)} = \sum_{\beta\in\textsc{MIs}} R^{(+)}_{\alpha\beta}(\gamma)\, I_{\beta},\qquad \textrm{for }\alpha\in\textrm{MIs}.
\end{equation}
By inverting $R^{(+)}$ and shifting $\gamma\to\gamma-1$ we obtain the recurrence relation we seek
\begin{equation}
  I_\alpha^{(\gamma-1)} = \sum_{\beta\in\textsc{MIs}} R^{(-)}_{\alpha\beta}(\gamma)\, I_{\beta},\qquad \textrm{for }\alpha\in\textrm{MIs}, \label{eq:recrel-}
\end{equation}
with
\begin{equation}
  R^{(-)}_{\alpha\beta}(\gamma) = \left[ R^{(+)}(\gamma-1)  \right]^{-1}_{\alpha\beta}.
\end{equation}
We thus reduce the second term in eq.~\eqref{eq:dIalpharec} to shifted MIs with exponent $\gamma-1$ and then use recurrence relation~\eqref{eq:recrel-} to rewrite the shifted MIs in terms of unshifted MIs.  We thus obtain a DE for the (unshifted) MIs of the original family, as in eq.~\eqref{eq:de}.

While the strategy we just outlined is viable, the appearance of shifted integrals in intermediate stages and the need of finding and using recurrence relations make it quite involved.  In the following, we present a method where shifted integrals do not appear at any stage of the calculation.

\subsection{Differential equations via differential operators}

A more elegant method of deriving differential equations, that better fits the general approach we follow in this paper, consists in deriving an operator $\hat O_x$ that realizes differentiation with respect to $x$.  In other words, we find an operator $\hat O_x$ of order $o$
\begin{equation}
  \hat O_x = c_0^{(x)}(\z) + \sum_j c_j^{(x)} (\z)\, \partial_j + \sum_{j_1\leq j_2}c_{j_1 j_2}^{(x)}(\z)\, \partial_{j_1}\partial_{j_2}+\cdots+\sum_{j_1\leq\cdots\leq j_o}c_{j_1\cdots j_o}^{(x)}(\z)\, \partial_{j_1}\cdots\partial_{j_o} \label{eq:Oo}
\end{equation}
with polynomials $c_{j_1 j_2\cdots}^{(x)}$, such that
\begin{equation}
  \hat O_x\, u(\z) = \partial_x\, u(\z). \label{eq:Ode}
\end{equation}
Once such an operator is found, since it contains only derivatives with respect to the integration variables, differentiation with respect to $x$ can be obtained by integrating by parts all derivatives in $\hat O_x$, namely
\begin{align}
  \partial_x I_\alpha ={}& \int (\hat O_x u)\, \varphi_\alpha + \int u\, (\partial_x \varphi_\alpha) \vphantom{+\sum_j} \nn
                     ={}& \int u\, (\varphi_{\alpha} c_0^{(x)}) - \sum_j \int u\, (\partial_j c_j^{(x)} \varphi_\alpha) +\cdots+(-1)^o\sum_{j_1\leq\cdots\leq j_o}\int u\, \partial_{j_1}\cdots\partial_{j_o}(c_{j_1\cdots j_o}^{(x)} \varphi_\alpha)\nn
                           & + \int u\, (\partial_x \varphi_\alpha). \label{eq:dIalpha}
\end{align}
Under our assumptions (see section~\ref{sec:def}), all integrals on the r.h.s.\ of the last equality belong to our integral family and can thus be reduced to MIs.  This reduction yields the DEs in eq.~\eqref{eq:de}.  This is our preferred method for computing DEs in the context of this paper.

It is worth observing that the operator $\hat O_x$ satisfying eq.~\eqref{eq:Ode} is not unique, but the difference between two operators satisfying eq.~\eqref{eq:Ode} is a parametric annihilator.  Hence, we can say that $\hat O_x$ is unique modulo the addition of parametric annihilators of $u$.

\section{Computing annihilators via linear constraints}
\label{sec:comp}

In this section we illustrate a method for computing parametric
annihilators up to a certain order and polynomial degree.  The method
has been implemented in the \textsc{CALICO} package.  It consists in
two steps.  The first translates the annihilator equation in
eq.~\eqref{eq:A} into a syzygy equation for the polynomials
$c_{j_1\cdots j_k}(\z)$ in eq.~\eqref{eq:Ao}.  The second step finds
syzygy solutions by constraining a polynomial ansatz.

\subsection{From annihilators to syzygies}

A syzygy equation is an equation of the form
\begin{equation}
  \f(\z) \cdot \g(\z) = 0 \label{eq:syz}
\end{equation}
where
\begin{equation}
 \f(\z) = \{f_1(\z),\ldots,f_n(\z)\}
\end{equation}
is a list of \emph{known polynomials} and
\begin{equation}
  \g(\z) = \{g_1(\z),\ldots,g_n(\z)\}
\end{equation}
is a list of \emph{unknown polynomials} to be found.  If $\g^{(k)}(\z)$ are a set of solutions of eq.~\eqref{eq:syz}, then
\begin{equation}
  \g(\z) = \sum_k\, p_k(\z)\, \g^{(k)}(\z), \label{eq:gensyzsol}
\end{equation}
where $p_k(\z)$ are arbitrary polynomials, is also a solution.  In the language of algebraic geometry, we say that syzygy solutions form a \emph{module}.  We generally wish to find a set of \emph{generators} $\{\g^{(k)}\z)\}_k$ that is minimal, i.e.\ that are independent of each other with respect to combinations of the form in eq.~\eqref{eq:gensyzsol}.  Finding a complete set of generators is however often unnecessary.  We can, indeed, generally limit ourselves to a subset of generators which is sufficient for the solution of a certain problem, for instance all independent generators up to a maximum degree.

One can easily use eq.~\eqref{eq:A} to identify polynomials
$c_{j_1,\ldots,j_k}(\z)$ of an annihilator of order $o$ (see
eq.~\eqref{eq:Ao}) as the unknown polynomials $\g(\z)$ of a suitable
syzygy equation.  We first rewrite eq.~\eqref{eq:A} as
\begin{equation}
  \frac{1}{u(\z)}\, \hat A\, u(\z) = 0. \label{eq:1ouAu}
\end{equation}
After inserting eq.~\eqref{eq:Ao}, with symbolic unknown polynomials
$c_{j_1,\ldots,j_k}(\z)$, into the previous equation and simplifying
out non-rational terms, we collect the resulting rational
expression on the l.h.s.\ under a common denominator.  By imposing the
vanishing of the numerator of such expression we obtain a syzygy
equation for the polynomials $c_{j_1,\ldots,j_k}(\z)$.

The strategy can be clarified by means of an explicit example.
Consider \emph{first-order annihilators} (i.e.\ with $o=1$) of a twist
having the form in eq.~\eqref{eq:u}, which is the most common in our applications.  Inserting eq.~\eqref{eq:Ao} into eq.~\eqref{eq:1ouAu} we obtain
\begin{equation}
  c_0(\z) + \sum_{j=1}^n c_j(\z)\, \sum_k \gamma_k \frac{\partial_j B_k(\z)}{B_k(\z)}=0,
\end{equation}
which is a rational equation.   After collecting under common denominator and setting the numerator to zero we obtain
\begin{equation}
  c_0(\z) \prod_{k} B_k(\z) + \sum_{j=1}^n c_j(\z)\, \sum_k \gamma_k \left(\partial_j B_k(\z) \right) \prod_{l\neq k} B_l(\z) =0,
\end{equation}
which has now the form of eq.~\eqref{eq:syz} by identifying
\begin{align}
  \f(\z) = {} & \Big\{ \prod_{k} B_k(\z), \sum_k \gamma_k \left(\partial_{z_1} B_k(\z) \right) \prod_{l\neq k} B_l(\z), \ldots, \sum_k \gamma_k \left(\partial_{z_n} B_k(\z) \right) \prod_{l\neq k} B_l(\z) \Big\} \nn
  \g(\z) ={}& \{ c_0(\z), c_1(\z), \ldots , c_n(\z) \}.
\end{align}
The same strategy can be applied to higher-order annihilators and
annihilators for twists having a different form, e.g.\ having an
exponential factor as well.

\subsection{Solving syzygy equations via linear constraints}

Syzygy equations are often solved via techniques of computational
algebraic geometry, such as Gr\"obner basis techniques.  However, they
can also easily be reduced into a problem of linear algebra by turning
them into linear constraints for the coefficients of a polynomial
ansatz. We prefer the latter strategy, as it allows us to exploit efficient sparse linear
solvers based on finite-field techniques and functional reconstruction
routines. Moreover, as already mentioned, we are generally interested
to find solutions up to a maximal degree, which in turn can be
easily achieved by writing an ansatz having this property.

The strategy of finding syzygy solutions via linear constraints is
well known and has already been used in the context of loop integrals (see e.g.~\cite{Schabinger:2011dz}).  Here, we give a brief review of it.
Given a syzygy equation of the form in eq.~\eqref{eq:syz}, we make an
ansatz for its solution\footnote{In our ansatz the coefficients $c_{j\alpha}$ are rational functions in the external parameters $x$ and regulators.  In principle, without loss of generality, one could also make a polynomial ansatz with respect to these additional parameters, in terms of rational unknown numerical coefficients (see e.g.~\cite{Bendle:2019csk}).  We however found that our strategy is generally more efficient when combined with finite-field techniques and functional reconstruction algorithms, besides being slightly easier to generalize to polynomial decomposition problems (see eq.~\eqref{eq:polydec}).}
\begin{equation}
  \g(\z) = \sum_{j}\, \sum_{\alpha}c_{j \alpha}\, \z^\alpha\, \hat e_j \label{eq:syzansatz}
\end{equation}
with $\hat e_j$ being the unit vector in the $j$-th direction.  The
second sum is over all the multidimensional exponents $\alpha$ such
that $|\alpha|\leq d$ for some maximum degree
$d$.

After inserting this ansatz into eq.~\eqref{eq:syz} we impose that the coefficient of each monomial in $\z$ on the l.h.s.\ of the equation vanishes.  This yields a linear system of equations for the coefficients $c_{j\alpha}$.  Each independent solution of this system yields a syzygy solution.  Since the linear system is homogeneous in the unknowns $c_{j\alpha}$, we can distinguish two cases.  The first is when there's no solution but the trivial one, $c_{j\alpha}=0$ for all $j,\alpha$.  In this case there's no non-trivial syzygy solution compatible with the ansatz.  In the second case, the system has a non-trivial solution which constraints a subset of the coefficients $c_{j\alpha}$ --- which we call \emph{dependent} coefficients --- to be specific linear combinations of a second subset --- which we call \emph{independent coefficients}.  More explicitly
\begin{equation}
  c_{j \alpha} = \sum_{k,\beta} a_{j k \alpha \beta}\, c_{k \beta}, \label{eq:syzcsol}
\end{equation}
where sum over $k$ and $\beta$ on the r.h.s.\ runs over the pairs of indexes and exponents which identify the independent coefficients $c_{k \beta}$, while $a_{j k \alpha \beta}$ are rational functions of the free parameters of the problem.

Hence, we obtain a list of syzygy solutions by substituting eq.~\eqref{eq:syzcsol} into the ansatz~\eqref{eq:syzansatz} and setting all independent coefficients but one to zero and the remaining coefficient to an arbitrary non-zero value (typically 1), for all possible choices of the non-zero independent coefficient.

Since we are solving an underdetermined system, the list of independent coefficients and thus the explicit form of the solution depend on the order of the unknowns, namely on which unknowns are eliminated with higher priority. Unknown coefficients $c_{j\alpha}$ which compare lower are eliminated with lower priority and are thus preferred to be in the independent subset. Notice that, via the procedure described before, terms which correspond to independent coefficients appear less frequently in the syzygy solutions, since in all but one they are set to zero. There are two criteria that determine the order of the coefficients that we use. The first criterion is based on the \emph{monomial order}. In particular we have $c_{j\alpha}<c_{k\beta}$ if $\z^\alpha>\z^\beta$. This means that, monomials that compare higher generally appear less frequently in the syzgy solutions. At the time of writing, the \textsc{CALICO} package uses the \emph{degree reverse lexicographic order} to compare monomials. The second criterion is the \emph{position}, namely $c_{j\alpha}<c_{k\beta}$ if $j>k$, which means that terms in the rightmost positions of the solutions appear less frequently than terms in the leftmost positions, in the syzygy solutions. By default \textsc{CALICO} uses a \emph{term over position order}, meaning the first criterion takes higher priority, while the second is used as a tie-breaker. The opposite, namely the \emph{position over term order}, can also be optionally used.

The syzygy solutions found as we just described are not guaranteed to be independent modulo combinations of the form in eq.~\eqref{eq:gensyzsol}, except for the case where $p_j$ are all constants.  In order to obtain a minimal set of generators for the syzygy solutions, it is thus convenient to \emph{filter out} some of them, by restricting the ansatz in eq.~\eqref{eq:syzansatz}.  This is discussed in the next subsection.

\subsection{Filtering out  solutions}
\label{sec:filtering}

As explained in section~\ref{sec:ann}, we are interested in finding a minimal set of generators of parametric annihilators, from which other annihilators can be derived via monomial multiplication, differentiation with respect to the variables~$\z$ and combinations thereof.  Annihilators are, in turn, mapped into solutions of syzygy equations, which we find by turning them into linear constraints applied to the coefficients of an ansatz.  In order to remove, from these, solutions that are not independent upon monomial multiplication and differentiation, we generally want to further constrain the initial ansatz in~\eqref{eq:syzansatz}.

We generally find annihilators in multiple steps.  In each step, we compute annihilators of a specific order $o$ and degree $d$.  The latter is identified as the degree of the polynomials $c_{j_1 j_2\cdots}(\z)$ in $\z$.  We proceed from lower to higher orders $o$.  Furthermore, annihilators of the same order are computed from lower to higher degrees $d$ in the variables $\z$.  When finding annihilators of order $o$ and maximal degree $d$, we generally want to exclude solutions which can be found by
\begin{itemize}
\item differentiating annihilators of lower orders,
\item multiplying solutions of the same order but lower degree by a monomial $\z^\alpha$
\end{itemize}
and linear combinations of the above, since they are generated by solutions found in previous steps.  Before finding annihilators of order $o$ and degree $d$, we thus generate a list of annihilators of the same order and maximal degree generated by those already known, as we outlined in the two points above.  Each $o$-th order annihilator, via the method described in the previous subsection, can be mapped into a solution of a suitable syzygy equation.  After this mapping, it takes the form in eq.~\eqref{eq:syzansatz}, now with \emph{known} coefficients $c_{j\alpha}$.  Each syzygy solution is, in turn, in a one-to-one correspondence to a linear equation, namely
\begin{equation}
  \sum_{j,\alpha}c_{j \alpha}\, \z^\alpha\, \hat e_j\quad \leftrightarrow \quad \sum_{j,\alpha}c_{j \alpha}\, y_{\alpha j} = 0,
\end{equation}
where $y_{\alpha j}$ are the unknowns of this equation.  By solving the system of equations generated this way from all the known solutions of order $o$ and maximal degree $d$, for instance using Gaussian elimination, we are --- via the mapping above --- effectively taking linear combinations of the syzygy solutions.  For all dependent unknowns $y_{\alpha j}$ of the solutions, we thus set to zero the corresponding coefficient $c_{\alpha j}$ in the ansatz in eq.~\eqref{eq:syzansatz}.  This guarantees that the new syzygy solutions, found as described in the previous subsection, are independent of those which are already known.  The order we use for the unknowns $y_{j\alpha}$ in this system is reversed compared to the one described above for the $c_{j\alpha}$.

\subsection{Computing other differential operators}

Besides annihilators, in this paper we are also interested in other differential operators, such as those that are equivalent to differentiation with respect to external parameters, as explained in section~\ref{sec:de}.  Their calculation is similar to the one of annihilators, with some important differences that we quickly illustrate below.

Consider, for instance, the differential operator $\hat O_x$ defined in eq.~\eqref{eq:Ode}.  Similarly to the case of annihilators, we insert the general form of an operator of order $o$ given in eq.~\eqref{eq:Oo} into eq.~\eqref{eq:Ode}, divide by $u(\z)$ and put everything under common denominator to obtain a polynomial equation for the polynomial coefficients $c_{j_1\cdots j_k}^{(x)}(\z)$. This has a form that is similar to the one of a syzygy equation, except that it has a non-vanishing right-hand side
\begin{equation}
  \f(\z) \cdot \g(\z) = h(\z), \label{eq:polydec}
\end{equation}
where $\f(\z)$ and $\g(\z)$ are lists of known and unknown polynomials respectively, as above, while $h(\z)$ on the r.h.s.\ is also a known polynomial. Note that we generally need only one solution for this equation.  Any other solution can be obtained by adding to it a solution of the syzygy equation~\eqref{eq:syz}.

Once again, we clarify this by means of an example.  Consider a twist of the form in eq.~\eqref{eq:u}.  Following the steps outlined above for a first-order operator $\hat O_x$, we obtain
\begin{equation}
  c_0^{(x)}(\z) \prod_{k} B_k(\z) + \sum_{j=1}^n c_j^{(x)}(\z)\, \sum_k \gamma_k \left(\partial_j B_k(\z) \right) \prod_{l\neq k} B_l(\z) = \sum_k \gamma_k \left(\partial_x B_k(\z) \right) \prod_{l\neq k} B_l(\z),
\end{equation}
which has the form in~\eqref{eq:polydec}, identifying
\begin{align}
  \f(\z) = {} & \Big\{ \prod_{k} B_k(\z), \sum_k \gamma_k \left(\partial_{z_1} B_k(\z) \right) \prod_{l\neq k} B_l(\z), \ldots, \sum_k \gamma_k \left(\partial_{z_n} B_k(\z) \right) \prod_{l\neq k} B_l(\z) \Big\} \nn
  \g(\z) ={}& \{ c_0^{(x)}(\z), c_1(\z), \ldots , c_n^{(x)}(\z) \} \nn
  h(\z) = {}& \sum_k \gamma_k \left(\partial_x B_k(\z) \right) \prod_{l\neq k} B_l(\z).
\end{align}

Eq.~\eqref{eq:polydec} can be mapped into a problem of linear algebra, following a similar strategy to the one we described in the previous subsections. This has also been implemented in the \textsc{CALICO} package. We still make an ansatz of the form in eq.~\eqref{eq:syzansatz}.  If solutions of the corresponding syzygy equation~\eqref{eq:syz} are known, they can be used to constrain the ansatz as already described for syzygy equations. This is not strictly necessary, because only one solution is needed, but it can improve performance.  By inserting the ansatz into eq.~\eqref{eq:polydec} and matching the coefficients of each monomial $\z^\alpha$ appearing on either side of the equation, we obtain a linear system of identities for the coefficients $c_{j\alpha}$.  This time, however, the system is not homogeneous in the unknowns, hence it may have no solution.  If the system is impossible, then there's no solution compatible with the ansatz and we may proceed by making a more general ansatz with a higher maximum degree $d$ or searching for an higher-order form of the operator $\hat O_x$.  If there is at least one solution, we set to zero any independent coefficient $c_{j\alpha}$ that has not been constrained by the system of equations, thus obtaining the solution we seek.  Once one solution has been found, the algorithm successfully terminates.

\subsection{Finite-field methods and implementation details}
\label{sec:ffdetails}

We now provide additional details about an implementation of the methods we outlined, which we publish with this work as the \textsc{Mathematica} package \textsc{CALICO}, whose usage is described in more details in section~\ref{sec:calico}. \textsc{CALICO} relies on the \textsc{FiniteFlow}~\cite{Peraro:2016wsq,Peraro:2019svx} program for solving linear systems. The latter uses a numerical sparse solver over finite fields to efficiently solve the system for numerical values of the input parameters, combined with functional and rational reconstruction techniques to recover the full analytic form of the solution.

The strategy we described starts with an analytic preparation which converts equations~\eqref{eq:A} and~\eqref{eq:Ode} into polynomial equations of the form in~\eqref{eq:syz} and~\eqref{eq:polydec} respectively. This is done using a Computer Algebra System~(CAS), in our case \textsc{Mathematica}. The main non-trivial optimization we optionally make here is avoiding the use of the explicit expressions of the polynomials which define the twist until after the equation has been cast into a polynomial form (see the function \texttt{CATTwist} described in section~\ref{sec:calico}).

After eq.~\eqref{eq:syz} or~\eqref{eq:polydec} has been obtained for a given $o$-th order operator, we need to make an ansatz for $\g(\z)$, as in eq.~\eqref{eq:syzansatz} and generate a linear system for the unknown coefficients $c_{j\alpha}$ appearing in it. If implemented naively, this step can easily become a bottleneck due to the size of the expressions involved, especially when using a CAS such as \textsc{Mathematica} that is not optimized for handling large expressions. Communication between \textsc{Mathematica} and \textsc{FiniteFlow} can also cause significant overhead when many large expressions are present.  These issues are mitigated by exploiting the general form of the equations as well as the features available in \textsc{FiniteFlow}, such as the possibility of combing core algorithms into more complex ones by defining them via computational graphs.

The \emph{known} polynomials appearing in our equations can be written as
\begin{align}
  \f(\z) ={}& \sum_{j\alpha} b_{j\alpha}\, \z^\alpha\, \hat e_j \nn
  h(\z) ={}& \sum_{j\alpha} b_{0\alpha}\, \z^\alpha,
\end{align}
where $b_{j\alpha}$ are \emph{known} rational functions (typically polynomials) of the free parameters of the problem. We notice that, when these expressions are inserted in equations~\eqref{eq:syz} or~\eqref{eq:polydec}, different coefficients $b_{j\alpha}$ always multiply either different unknowns~$c_{j\alpha}$ or different monomials~in $\z$ which correspond to different equations, hence they never mix. Indeed, each entry of the matrix that defines the system of equations for the coefficients $c_{j\alpha}$ coincides with one of the coefficients $b_{j\alpha}$.  We thus exploit this by first loading the coefficients $b_{j\alpha}$ in a computational graph of \textsc{FiniteFlow}~(after removing duplicate entries) and then using this to define the system for the unknowns~$c_{j\alpha}$ via list manipulations, whose performance is vastly superior to algebraic manipulations. These manipulations are meant to identify which coefficient $b_{j\alpha}$ corresponds to each entry of the matrix which defines the system, without performing any symbolic algebra. Moreover, the same set of coefficients~$b_{j\alpha}$ is reused for all choices of the maximum degree of the ansatz for $\g(\z)$. This greatly increases the performance, especially when many free parameters or many integration variables are present.

In our implementation, the maximum order and degree of annihilators to be found is specified by the user.  In realistic applications, one would need to adjust these until either no more annihilator is found or one is convinced that the ones which have been found are sufficient for the reduction to master integrals (this is analogous to how one normally chooses the seeds).

Finally, the filtering described in~\ref{sec:filtering} is performed numerically, i.e.\ by replacing all free parameters with arbitrary numerical values while solving the corresponding system, since the only information we need from it is a list of coefficients and corresponding monomials to exclude from the ansatz.

\section{Hypergeometric functions}

Families of integrals in the general form we defined in section~\ref{sec:def} are found in all sorts of applications.  They include classes of special functions, integrals representing expectations values of operators in quantum mechanics, correlation functions in quantum field theory, and loop integrals (see e.g.\ ref.~\cite{Cacciatori:2022mbi}).  In this section we present some application of the formalism we implemented to hypergeometric functions.

\subsection*{Function ${}_2F_1$}
As a simple application, consider the family of univariate integrals
\begin{align}
  I_\alpha ={}& \int_0^1 \dd z\, \varphi_\alpha(z)\, u(z) \nn
 \varphi_\alpha(z) = {}& z^\alpha,\qquad
  u(z)={}z^{b_2-1}\, (1-z)^{b_3-b_2-1}\, (1-x\, z)^{-b_1}, \label{eq:2F1fam}
\end{align}
which is related to the hypergeometric function ${}_2F_1$ by
\begin{equation}
  I_\alpha = \frac{\Gamma(b_2+\alpha)\Gamma(b_3-b_2)}{\Gamma(b_3+\alpha)}\, {}_2F_1(b_1,\, b_2+\alpha,\, b_3+\alpha;\, x).
\end{equation}
The parameters $b_i$ are analytic regulators, while $x$ is a free
parameter.  Since the integral is univariate, $\alpha$ here is an
integer exponent rather than a list of exponents.

We first find annihilators of the twist $u(z)$.  Using the algorithm
described above, we find that first order annihilators are generated by
\begin{equation}
    \hat A ={} c_0(z) + c_1(z)\, \partial_z
\end{equation}
with
\begin{align}
   c_0(z) ={}& 1 - b_2 + (b_3 - 2 + (b_2 - b_1 -1)\, x)\, z + (2 + b_1 - b_3)\, x\, z^2\nn
 c_1(z) ={} &  z - (1 + x)\, z^2 + x\, z^3 .
\end{align}
Using this annihilator, eq.~\eqref{eq:Atmpeq} yields
\begin{equation}
  -(\alpha + b_2) I_\alpha + (\alpha + b_3 + (1 + \alpha - b_1 + b_2) x) I_{\alpha+1} + (b_1 - b_3 -1 - \alpha)\, x\, I_{\alpha+2} = 0,
\end{equation}
which can be solved to express, say, $I_{\alpha+2}$ in terms of
$I_{\alpha+1}$ and $I_{\alpha}$, or alternatively $I_{\alpha}$ in
terms of $I_{\alpha+1}$ and $I_{\alpha+2}$.  By using this relation
for several values of $\alpha$, it is straightforward to see that this
family has two MIs, which we can choose as $I_0$ and $I_1$.  As an
example of such a reduction identity, we have
\begin{equation}
  I_2 = \frac{b_2}{(b_1 - b_3 - 1)\, x}\, I_0
  + \frac{(b_1 - b_2 - 1)\, x - b_3}{(b_1 - b_3 - 1)\, x}\, I_1.\label{eq:I2hypred}
\end{equation}
The integrals depend on the free parameter $x$ and we can derive
differential equations with respect to it.  In order to do that, as
explained in section~\ref{sec:de}, we first compute the
corresponding differential operator $\hat O_x$, which we find to be
\begin{equation}
  \hat O_x = c_0^{(x)}(z) + c_1^{(x)}(z)\, \partial_z
\end{equation}
with
\begin{align}
  c_0^{(x)}(z) ={}& \frac{(b_2-1) + (2 + b_1 - b_3)\, z}{1 - x} \nn
  c_1^{(x)}(z) ={}& \frac{z^2-z}{1 - x}.
\end{align}
Using eq.~\eqref{eq:dIalpha} we thus obtain
\begin{equation}
\partial_x I_\alpha = \frac{\alpha + b_2}{1-x}\, I_\alpha + \frac{b_1-b_3-\alpha}{1-x}\, I_{\alpha+1}. \label{eq:hypdx}
\end{equation}
Applying this to both MIs $I_0$, $I_1$ and using
eq.~\eqref{eq:I2hypred} to reduce $I_2$ to MIs, we obtain a system of
differential equations satisfied by the MIs
\begin{equation}
  \partial_x \begin{pmatrix} I_1 \\ I_0 \end{pmatrix} =
  \begin{pmatrix}
    \frac{b_1\, x - b_3}{(1-x)\, x} & \frac{b_2}{(1-x)\, x} \\
    \frac{b_1-b_3}{1-x} & \frac{b_2}{1-x}
  \end{pmatrix}\cdot \begin{pmatrix} I_1 \\ I_0 \end{pmatrix}.\label{eq:hypDEmat}
\end{equation}
Alternatively, we can use this relation to treat $\partial_x I_0$ as a
new master integral which replaces $I_1$.  This change of basis yields
\footnote{Given a system of equations
  $\partial_x G_i = \sum_{j}M_{ij}\,G_j$ and a new basis $\tilde G_i=\sum_j T_{ij}\, G_j$, the latter satisfies the new differential equation $\partial_x \tilde G_i = \sum_{j}\tilde M_{ij}\,\tilde G_j$ with $\tilde M = (\partial_x T)\cdot T^{-1} + T\cdot M \cdot T^{-1}$.  In our example, the new basis is $(\partial_x I_0, I_0)$ where $\partial_x I_0$ can be read from eq.~\eqref{eq:hypDEmat}.  The new first order DE satisfied by $(\partial_x I_0, I_0)$ is equivalent to a second-order DE satisfied by $I_0$.}
a second-order differential equation satisfied by $I_0$, namely
\begin{equation}
  x\, (1-x)\, \, \frac{\partial^2}{\partial {x^2}}\, I_0 = \Big((b_1+b_2+1)\, x - b_3\Big)\, \partial_x\, I_0 + \Big(b_1\, b_2\Big)\, I_0,
\end{equation}
which is the well-known differential equation satisfied by the
hypergeometric function ${}_2F_1(b_1,b_2,b_3;x)$.

\subsection*{Functions ${}_{n+1}F_n$}
As a generalization of the previous example, we consider the integral family defined as in eq.~\eqref{eq:Ialpha} with
\begin{equation}
  \varphi_\alpha(\z) = \z^\alpha, \qquad u(\z) = \Big( 1-x\, \prod_{j=1}^n z_j \Big)^{-a_{n+1}}\, \prod_{j=1}^n\left[ z_j^{a_j-1} (1-z_j)^{b_j-a_j-1} \right],
\end{equation}
where the integration contour is $z_j\in (0,1)$ for $j=1,\ldots,n$, while $a_j$ and $b_j$ are analytic regulators and $x$ a free parameter.

These integrals are related to the hypergeometric functions ${}_{n+1}F_n$ by
\begin{align}
  I_\alpha = {}&\prod_{j=1}\left[ \frac{\Gamma(a_j+\alpha_j)\Gamma(b_j-a_j)}{\Gamma(b_j+\alpha_j)} \right]\nn & \times {}_{n+1}F_n(a_1+\alpha_1,\ldots,a_n+\alpha_n,a_{n+1};b_1+\alpha_1,\ldots,b_n+\alpha_n; x).
\end{align}
We empirically find that, for various choices of $n$, one or more first-order parametric annihilators of $u(\z)$ of degree 3 and $n+2$ exist.  Higher-order generators of annihilators also exist for $n>1$ but, even though their generators are independent, the identities they generate are not independent of the ones obtained from first-order annihilators --- as we empirically verified they do not add independent constraints among the integrals.

Using eq.~\eqref{eq:Atmpeq} we generate identities satisfied by the integrals $I_\alpha$.  The solution of the linear system yields $n+1$ independent master integrals, which we may choose as
\begin{equation}
  \{I_{0\cdots0}\} \cup \{I_{\hat e_j}\}_{j=1}^n.
\end{equation}
We are also able to find a first-order differential operator $\hat O_x$ of polynomial degree~$n+1$ in $\z$ which implements differentiation with respect to the parameter $x$, defined as in equations~\eqref{eq:Oo} and~\eqref{eq:Ode}.  With this, we find differential equations for the $n+1$ MIs.  These identities have been checked against numerical evaluations of the hypergeometric functions.  Similarly to the previous case, this can also be cast as a $(n+1)$-th order differential equation for the master integral $I_{0\cdots 0}$.

\section{Loop integrals}
\label{sec:loop}

As already stated, the formalism we described can be applied to a wide variety of problems. The main focus of our work, however, is the study of loop integrals in dimensional regularization.  These are commonly organized into families. A family of loop integrals includes linear combinations of integrals that, in momentum space, have the form
\begin{equation}\label{eq:loopint}
   J_\alpha = J_{\alpha_1\cdots\alpha_n} = \int \prod_{j=1}^{\ell} \frac{\dd^d k_j}{i \pi^{d/2}}\,  \frac{1}{D_1^{\alpha_1}\dots D_n^{\alpha_n}},
 \end{equation}
for any multi-index of integers $\alpha$.  The denominators $D_j$ in the integrands are functions of the $\ell$ loop momenta $k_1, \dots, k_{\ell}$ and the $e$ linearly independent external momenta $p_1, \dots, p_e$.  These integrals generally contribute to scattering matrix elements or Green functions with $e+1$ external momenta, one of which is fixed as a linear combination of the others by momentum conservation.  The integral is performed in a generic number $d$ of dimensions and it is an analytic function of $d$.

Each loop integral family is thus defined by the set of generalized denominators $D_j$.  For each integral $J_\alpha$, we split these into two disjoint subsets: ~\emph{proper denominators} are $D_j$ with $\alpha_j > 0$, while \emph{irreducible scalar products (ISPs)} are $D_j$ with $\alpha_j \leq 0$ (i.e.\ contributing to the numerator of the integrand).
The denominators $D_j$ typically have the quadratic form
\begin{equation}
  D_j = l_j^2 - m_j^2, \label{eq:loopden}
\end{equation}
but one can also use the bi-linear form
\begin{equation}
  D_j = l_j \cdot v_j - m_j^2, \label{eq:loopdenbl}
\end{equation}
where $m_j$ are internal masses, $l_j$ are linear combinations of loop momenta and external momenta, while $v_j$ are linear combinations of external momenta only.
For loop integrals appearing in scattering amplitudes or Green functions in Quantum Field Theory~(QFT), the bi-linear form is only allowed for ISPs.

When computing loop integrals or deriving linear relations for them, it is useful to partition them into subsets called \emph{sectors}.  We define sectors
\begin{equation}
  S_\alpha = S_{\alpha_1\cdots\alpha_n}, \qquad \textrm{with } \alpha_j=0,1
\end{equation}
such that
\begin{equation}
  J_\alpha\in S_{\Theta(\alpha_1-1/2)\cdots \Theta(\alpha_n-1/2)}
\end{equation}
with $\Theta$ being the Heaviside step function.  Each sector can also be identified by its \emph{corner integral}, that is the only integral $J_\alpha$ of the sector with $\alpha_j\in\{0,1\}$ for all $j$.  Given two sectors $S_\alpha$ and $S_\beta$, we say that $S_\alpha$ is a \emph{subsector} of $S_\beta$ if $\alpha\neq \beta$ and $\alpha_j\leq \beta_j$ for all $j$.

It is also common to define, for each integral $J_\alpha$, the number
\begin{equation}
  t = \sum_{j=1}^n \Theta(\alpha_j-1/2),
\end{equation}
which is the same for integrals belonging to the same sector, as well as the degree or \emph{rank} $s$ of the numerator
\begin{equation}
  s = -\sum_{j=1}^n \alpha_j \Theta(-\alpha_j)
\end{equation}
and the numbers of \emph{dots}, that is the sum of the powers of denominators in excess with respect to those of the corner integral of the same sector, namely
\begin{equation}
  u = \sum_{j=1}^n (\alpha_j - 1) \Theta(\alpha_j-1/2).
\end{equation}
These numbers are often used to characterize the complexity of an integral.

In an integral family, we identify one or more \emph{top sectors}. These are a minimal set of sectors such that all integrals of interest, for solving a particular problem, either belong to those sectors or to their subsectors. In the examples illustrated in this paper, we focus on families with just one top sector.

Linear identities among loop integrals are often found using IBPs in momentum space~\cite{Tkachov:1981wb,Chetyrkin:1981qh}, which read
\begin{equation}
   \int \prod_{j=1}^{\ell} \frac{\dd^d k_j}{i \pi^{d/2}}\, \frac{\partial}{\partial k_m^\mu} \, \frac{v^\mu}{D_1^{\alpha_1}\dots D_n^{\alpha_n}}=0, \quad \textrm{with } v^\mu = k_i^\mu,p_i^\mu.
 \end{equation}
By carrying out the derivatives explicitly, we obtain a non-trivial linear combination of integrals that vanish.  These integrals, however, generally belong to the same family in eq.~\eqref{eq:loopint} only if we are able to rewrite all scalar products of the form $k_i\cdot k_j$ and $k_i\cdot p_j$ as linear combinations of generalized denominators.  This implies that the number $n$ of denominators must be
\begin{equation}
  n = \ell e + \frac{\ell (\ell +1)}{2}, \label{eq:fulln}
\end{equation}
which we can achieve by defining a suitable number of ISPs and adding them to the list of denominators that appear in the loop propagators of the amplitude or Green function that is being computed.  In the applications we will describe later, however, we will see that, when using parametric representations of loop integrals combined with annihilators, we are able to find integral identities without having to introduce a full set of ISPs.  This can drastically simplify the problem in some applications, by allowing a number $n$ of generalized denominators that is \emph{lower} that the one in eq.~\eqref{eq:fulln}.

\section{Baikov representations}

\subsection{Standard Baikov representation}
\label{sec:baikov}
The first parametric representation of loop integrals we discuss is the Baikov representation~\cite{Baikov:1996iu}, where loop integrals take the form in eq.~\eqref{eq:Ialpha}, namely
\begin{equation}
  J_\alpha = K\, I_\alpha, \label{eq:JtoIbaikov}
\end{equation}
with
\begin{equation}
  I_\alpha = \int \dd^n \z\, \frac{1}{\z^\alpha}\, B(\z)^\gamma \label{eq:baikov}
\end{equation}
with $\gamma=\frac{d-\ell -e - 1}{2}$.  This is equivalent to the identifications
\begin{equation}
  \varphi_\alpha = \frac{1}{\z^\alpha}, \qquad u(\z) = B(\z)^\gamma
\end{equation}
in eq.~\eqref{eq:Ialpha}.  The prefactor $K$ is the same for all integrals within a family, hence completely irrelevant when finding linear relations.  However, it generally contains a factor $B_0^{\gamma_0}$ --- with $\gamma_0$ a function of $d$ and $B_0$ independent of $\z$ but dependent on the external kinematic variables --- which must be taken into account when deriving DEs.  The integration contour is not important for the purposes of this paper, except for the property that boundary terms in IBPs can be set to zero.

The standard Baikov representation requires a full set of ISPs, such that the number $n$ of generalized denominators, equal to the number of integration variables in eq.~\eqref{eq:Ialpha}, is given by eq.~\eqref{eq:fulln}. Identities between integrals in the Baikov representation have already been extensively discussed in the literature, mostly using syzygy approaches, which we showed in section~\ref{sec:compsyz} to be equivalent to using identities generated by first-order annihilators.  Since a complete set of syzygy solutions for this parametrization is known in closed form~\cite{Bohm:2017qme}, we conclude that first-order annihilators (just like syzgy solutions) are generated by $n+1$ generators, all of which have degree 1.  We are not aware of examples where higher-order annihilators are needed for a complete reduction to master integrals in this representation.

We also recall that, as already stated in section~\ref{sec:compsyz}, additional syzgy equations or constraints are often used to exclude integrals with higher power of denominators from the identities. This strategy can be applied within both approaches but its discussion is outside the purposes of this work.

\subsection{Loop-by-loop Baikov representation}
\label{sec:lblbaikov}
A closely related representation of loop integrals is the \emph{loop-by-loop Baikov parametrization}~\cite{Frellesvig:2017aai}. This consists in applying the Baikov parametrization one loop at the time.  The two main differences with the standard Baikov parametrization are:
\begin{itemize}
\item the twist $u(\z)$, up to a common overall factor, takes the form in eq.~\eqref{eq:u}, more precisely it is the product of $2\ell-1$ polynomials $B_j(\z)$ raised to exponents $\gamma_j$ that depend on the number of space-time dimensions $d$;
\item the number $n$ of integration variables is generally lower than the one in eq.~\eqref{eq:fulln}, hence fewer ISPs are generally needed to define a family.
\end{itemize}

Via the method of annihilators, we can derive identities in the loop-by-loop Baikov parametrization of loop integrals, thus having fewer ISPs and integration variables with respect to the standard Baikov parametrization or the momentum representation.

\subsubsection*{Example}

\begin{figure}
  \centering
  \includegraphics[width=0.35\textwidth]{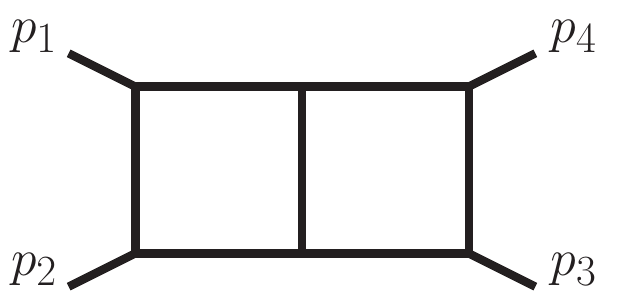}
  \caption{Double-box.}
  \label{fig:dbox}
\end{figure}

We consider a massless double-box integral family.  Its top sector has 7 proper denominators and corresponds to the diagram depicted\footnote{The conventions for the graphs depicted in this paper are: black single lines correspond to massless particles (on-shell if external), external double lines are off-shell external momenta, coloured lines correspond to massive particles (having the same mass, unless explicitly stated otherwise).} in fig.~\ref{fig:dbox}.
When IBPs in momentum representation are used, 2 additional generalized denominators are needed, to match eq.~\eqref{eq:fulln}.  In the loop-by-loop Baikov representation, instead, only 1 additional generalized denominator is required.  The full list of $n=8$ generalized denominators can be chosen as in eq.~\eqref{eq:loopden} with $m_j=0$ and
\begin{align}
& l_1 = k_1, \quad && l_2 = k_1+p_1, \quad && l_3 = k_1+p_1+p_2,\nn
& l_4 = k_1+k_2, \quad && l_5 = k_2, \quad && l_6 = k_2-p_1-p_2-p_3,\nn
& l_7 = k_2-p_1-p_2, \quad && l_8 = k_1+p_1+p_2+p_3.
\end{align}
For simplicity, we consider massless external legs, $p_j^2=0$. This implies there are two independent invariants, which we may choose as
\begin{equation}
  s_{12} = (p_1+p_2)^2, \qquad s_{23} = (p_2+p_3)^2.
\end{equation}
We consider integrals of the form in eq.~\eqref{eq:loopint} with
\begin{equation}
  \alpha_j\leq 0,\qquad \textrm{for }j>7.
\end{equation}

By integrating out the $k_2$-dependent loop first and $k_1$ second,\footnote{We may also chose to integrate $k_1$ first, but in this case the generalized denominator $D_8$ needs to be replaced with $D_9$, with $l_9=k_2-p_1$. Note that both $D_8$ and $D_9$ are instead needed for representations satisfying eq.~\eqref{eq:fulln}, i.e.\ requiring a total of 9 generalized denominators for our example.} the loop-by-loop Baikov representation yields a twist of the form
\begin{equation}
  u(\z) = C(d)\, B_0^{\frac{4-d}{2}}\, B_1(\z)^{\frac{d-5}{2}}\, B_2(\z)^{\frac{4-d}{2}}\, B_3(\z)^{\frac{d-5}{2}},\label{eq:lbluexample}
\end{equation}
where $B_0$ is independent of $\z$, hence it is irrelevant when finding linear relations, but it depends on the invariants $s_{ij}$, therefore it needs to be taken into account for deriving differential equations.  The explicit expressions of $B_j(\z)$ are reported in the examples within the public repository of the \textsc{CALICO} package.  The factor $C(d)$, instead, only depends on $d$ and is thus irrelevant for both applications.

Using \textsc{CALICO}, we find 8 first-order annihilators of degree 1 and 1 first-order annihilator of degree 2. We find that these, combined with some symmetry relations, are sufficient for a complete reduction to master integrals. We also empirically found that this twist admits 9 independent second-order annihilators of degree 1, which however are not required for the reduction to master integrals.

For this example, we used symmetry relations found with the help of the private package \textsc{FFIntRed}. For this purpose, we used the momentum representation of this integral family and searched for shifts of loop momenta which map generalized denominators into each other -- modulo permutations of external momenta which do not change the two invariants $s_{12}$ and $s_{23}$. It was sufficient to generate these symmetries up to rank $s=1$. These supplementary identities generally introduce an additional scalar product $k_2\cdot p_1$ which is not a linear combination of the 8 generalized denominators. This is, however, easily removed by taking suitable linear combinations of symmetry relations.  We thus simply generated symmetry relations in momentum space involving integrals with rank $s=0,1$, and used Gauss elimination to remove from them integrals having the unwanted scalar product $k_2\cdot p_1$.

The reduction yields 8 independent master integrals, in agreement with the traditional Laporta algorithm in momentum space. A possible choice of them is
\begin{equation}
  \{I_{1111111 -1}, I_{11111110}, I_{11010110}, I_{11110100},I_{10101010}, I_{10110100}, I_{01010100}, I_{10010010}\}.
\end{equation}

We also found differential operators $\hat O_{s_{12}}$ and $\hat O_{s_{23}}$ as first-order operators of degree 1. Using this, we take derivatives of the master integrals (cfr.\ eq.~\eqref{eq:dIalpha}) which, combined with the linear relations above, yield a system of differential equations for the master integrals.

\section{Duals of loop integrals}
\label{sec:duals}
In this section, we apply the formalism of parametric annihilators to
\emph{duals of loop integrals}.  Duals of loop integrals play a
crucial role within the framework of \emph{intersection theory}.  The
latter defines scalar products, called \emph{intersection numbers},
between integrals having a form analogous to the one in
eq.~\eqref{eq:Ialpha}, including --- as we have also reviewed in this
paper --- various parametrizations of loop integrals.  A review of
intersection theory is outside the scope of this paper.  We refer
the reader to~\cite{Mizera:2017rqa,Mizera:2019gea,Mastrolia:2018uzb,Frellesvig:2019uqt,Weinzierl:2020xyy,Fontana:2022yux} and references therein.
In the following, after some motivation, we recall only the main details that are relevant to our application of parametric
annihilators to duals of loop integrals.

The definition of intersection numbers, similarly to any other scalar
product, relies on the one of a \emph{dual} vector space.  While dual integrals are often studied in the context of intersection theory, there are several reasons to appreciate an alternative method to compute the linear relations they satisfy.
We give a quick review of some of them. Modern methods for computing intersection numbers involve deriving differential equations satisfied by dual integrals in fewer integration variables.
While these differential equations can be (and usually are) derived using intersection numbers, our approach offers an alternative to that.  Even when such differential equations are computed via intersection numbers, one can still avoid the appearance of shifted integrals by expressing derivatives via differential operators as in section~\ref{sec:de}, potentially simplifying their calculation.  Moreover, once the \emph{metric}, i.e.\ a set intersection numbers between MIs and dual MIs, has been computed --- possibly choosing the MIs in a way that makes this calculation as easy as possible --- any other intersection number is uniquely fixed by reductions of regular integrals and dual integrals. While the intersection numbers themselves are often used for reductions, they are also interesting mathematical objects by themselves and can give important insights about certain classes of integrals (see e.g.~\cite{Duhr:2023bku,Duhr:2024rxe,Duhr:2024xsy}).  Hence, having an alternative method for computing them, after the metric has been fixed, can be highly beneficial.  Finally, this method enables many checks, including consistency checks between intersection numbers and reduction identities of dual integrals.

We now briefly review how dual integrals can be defined, following the approach of ref.~\cite{Fontana:2023amt}.  For properly
regulated integrals, the dual space can be defined by replacing
$u(\z) \to 1/u(\z)$ in eq.~\eqref{eq:Ialpha} --- which is equivalent to
$\gamma\to -\gamma$ in the Baikov parametrization~\eqref{eq:baikov}.
Unfortunately this is problematic for loop integrals, since it
yields impossible systems of equations to be solved during the
calculation of intersection numbers.  When using the Baikov
parametrization, this is due to singularities at $z_j\to 0$ in $\varphi_\alpha(\z)$ that are
not properly regulated by the twist $u(\z)$.

In the following, up to a relabeling of the integration variables, we
will assume that $(z_1,\ldots,z_m)$ are proper denominators of the top
sector, while $(z_{m+1},\ldots,z_n)$ are its ISPs. This implies that
$\alpha_j\leq 0$ for $j>m$.  A possible solution~\cite{Frellesvig:2019uqt} to the issue above consists
in replacing the twist $u(\z)$ with
\begin{equation}
  u(\z) \to u_\rho(\z) = u(\z)\, \z^{\rho},
\end{equation}
where
\begin{equation}
  \rho = (\rho_1,\ldots,\rho_m,0,\ldots,0) \label{eq:rho}
\end{equation}
is a list of additional regulators for the proper
denominators.  While intersection numbers are singular in the limit
$\rho_j\to 0$, the coefficients of a reduction to MIs are finite.
However, since this limit can only be taken at the very end, the need
of additional regulators constitutes a major bottleneck and drawback
of this approach.

In references~\cite{Caron-Huot:2021xqj,Fontana:2023amt}, different approaches to defining the space of duals
of loop integrals were presented.  These sidestep the need of
regulators.  Here we focus on the one proposed in~\cite{Fontana:2023amt}, which
exploits our freedom of choice of dual loop integrals to
effectively avoid the need of performing algebraic operations using regulators.  More precisely, it consists
in choosing dual integrals $I_\alpha$ of the form
\begin{equation}
  I_\alpha = \prod_{j=1}^m \rho_j^{\Theta(\alpha_j-1/2)} \int \dd^n \z \, \frac{1}{\z^{\alpha-\rho}} \, B(\z)^{-\gamma}.
\end{equation}
with $\rho$ defined in~\eqref{eq:rho} and systematically work on the leading coefficients of the $\rho_j\to 0$ limit that yield a finite contribution to intersection numbers (see below for more details).  In other words, dual integrals are multiplied by a factor $\rho_j$ if $\alpha_j>0$.  Ref.~\cite{Brunello:2023rpq} shows (as a formal proof in the univariate case and heuristically in the multivariate one) that the approaches of references~\cite{Caron-Huot:2021xqj} and~\cite{Fontana:2023amt} yield equivalent results for intersection numbers, despite being significantly different in their formulation and computationally.

While, in the context of intersection theory, the factors $z_j^{\rho_j}$ are commonly absorbed into the definition of $u(\z)$, for the purposes of this paper it is more convenient to define
\begin{equation}
  \label{eq:phidual}
  \varphi_\alpha(\z) = \prod_{j=1}^m \rho_j^{\Theta(\alpha_j-1/2)} \, \frac{1}{\z^{\alpha-\rho}}, \qquad u(\z) = B(\z)^{-\gamma}
\end{equation}
for duals of loop integrals in the Baikov parametrization.  We can similarly define duals of integrals in the loop-by-loop Baikov parametrization, with the only difference that $u(\z)$ will take the form in eq.~\eqref{eq:u} with opposite exponents $\gamma_j\to -\gamma_j$ with respect to loop integrals.  When computing intersection numbers, if $\varphi_\alpha$ is such that $\alpha_j>0$, then intersection numbers involving the dual integral $I_\alpha$ have a $1/\rho_j$ pole that simplifies their $\rho_j$ prefactor yielding a finite result.  Hence, when $\rho_j\to 0$, we can ``effectively'' regard a factor $z_j^{-\alpha_j}$ of the integrand as being $\O(1/\rho_j)$ if $\alpha_j>0$ and  $\O(1)$ if $\alpha_j\leq 0$.  Our prescription of working on the leading terms of the $\rho_j\to 0$ limit that yield finite contributions to intersection numbers effectively amounts to the following rules
\begin{equation}
  \rho_j^2\, \frac{1}{\z^{\alpha-\rho}}\, u(\z) \to 0 \quad \forall \alpha, \qquad\quad \rho_j\, \frac{1}{\z^{\alpha-\rho}}\, u(\z)\to 0 \quad \textrm{if }\alpha_j\leq 0,\label{eq:rhorules}
\end{equation}
which can be implemented as substitutions at the integrand level when dealing with dual integrals.  For a more comprehensive justification of these rules we refer to section~4 of~\cite{Fontana:2023amt}.

\subsection{Reduction of dual integrals via annihilators}

Applying the annihilator approach to dual integrals is straightforward.  First, annihilators of $u(\z)=B(\z)^{-\gamma}$ can obviously be computed using the methods described in the previous sections.  In practice, one can use the same annihilators of loop integrals in the Baikov representation and simply replace $\gamma\to -\gamma$ (or, equivalently, $d\to 2 + 2\ell + 2 e - d$) in their expression.  Once a set of annihilators is known, identities are found, as before, applying eq.~\eqref{eq:Atmpeq}.  The latter requires knowing the behaviour of $\varphi_\alpha$ under multiplication by monomials and differentiation.  Our rules in eq.~\eqref{eq:rhorules} imply that, for monomial multiplication, with an exponent $\beta_j\geq 0$,
\begin{equation}
  z_j^{\beta_j}\, \varphi_\alpha(\z) = \begin{cases} \varphi_{\alpha-\beta_j \hat e_j}(\z)\quad & \textrm{if } \alpha_j>\beta_j \textrm{ or } \alpha_j\leq 0 \\
  0 \quad& \textrm{if } 0 < \alpha_j\leq\beta_j. \end{cases} \label{eq:dualmon}
\end{equation}
This shows that $\varphi_\alpha$ \emph{cuts} the propagators $j$ such that $\alpha_j>0$.  Differentiation instead works with the following rules:
\begin{equation}
  \partial_j\, \varphi_\alpha(\z) =  \begin{cases} -(\alpha_j - \delta_{\alpha_j 0})\, \varphi_{\alpha+\hat e_j}(\z) \quad & \textrm{if } j\leq m\\
    -\alpha_j\, \varphi_{\alpha+\hat e_j}(\z) \quad & \textrm{if } j> m. \end{cases} \label{eq:dualdiff}
\end{equation}
Hence derivatives follow the normal rules (as if regulators $\rho$ were not present) except for the case of zero exponents of regulated integration variables.  If $\alpha_j=0$ for $j\leq m$, then $\partial_j$ generates the denominator $1/z_j$.

We stress that, in practice, we only need to implement the rules in eq.~\eqref{eq:dualmon} and~\eqref{eq:dualdiff}, hence regulators never explicitly appear in our calculation.  Of course, besides reduction identities, one can also derive differential equations for dual integrals, following section~\ref{sec:de}.

From equations~\eqref{eq:dualmon} and~\eqref{eq:dualdiff} it follows that, as already noted in the literature~\cite{Fontana:2023amt}, reduction identities and differential equations for dual integrals have a block triangular structure that is transposed with respect to the one of loop integrals. For this reason, we find that a good choice of weight for duals of loop integrals would assign a lower weight to integrals with higher values of $t$, i.e.\ with more proper denominators, which is the opposite of what is commonly done for loop integrals.

We tested this approach on several examples and successfully compared
against reductions performed using intersection numbers, with the
algorithm of ref.~\cite{Fontana:2023amt}.

\subsection{Computation of connection matrices}
An important ingredient for computing multivariate intersection numbers using the recursive method in~\cite{Mizera:2019gea,Frellesvig:2019uqt} are the so-called connection matrices.  We rewrite eq.~\eqref{eq:Ialpha} as
\begin{equation}
  \label{eq:Ialpharec}
  I_{\alpha} = \int \dd z_{k+1}\cdots \dd z_{n}\, I_\alpha^{(k)}
\end{equation}
with
\begin{equation}
  I_\alpha^{(k)} \equiv \int \dd z_1\cdots \dd z_{k}\,  \varphi_{\alpha}(\z)\, u(\z).
\end{equation}
The $k$-fold integral $I_{\alpha}^{(k)}$ has a parametric dependence on
$z_{k+1},\ldots,z_{n}$. Hence, one can find a basis of master integrals for $I_{\alpha}^{(k)}$ and derive differential equation with respect to $z_{k+1}$
\begin{equation}
  \label{eq:inde}
  \frac{\dd}{\dd z_{k+1}}I^{(k)}_\alpha = \sum_{\beta \in \textrm{MIs}}\Omega^{(k)}_{\alpha \beta}\, I^{(k)}_\beta.
\end{equation}
The calculation of the \emph{connection matrix} $\Omega^{(k)}_{\alpha \beta}$ for \emph{duals} of Feynman integrals, with $k=1,\ldots,n-1$, is a key ingredient of the recursive method for computing $(k+1)$-variate intersection numbers in terms of $k$-variate intersection numbers.

The connection matrices can be computed with the formalism described in this paper, simply by treating $z_{k+1},\ldots,z_{n}$ as free parameters, hence following section~\ref{sec:de} to compute differential equation with respect to them. In particular, following this strategy, differential operators $\hat O_{z_{k+1}}$ are written in terms of derivatives w.r.t.\ the first $k$ integration variables, which are thus turned into linear combinations of dual integrals using eq.~\eqref{eq:dualmon} and~\eqref{eq:dualdiff}.

\subsubsection*{A simple example}
\begin{figure}
  \centering
  \includegraphics[width=5cm]{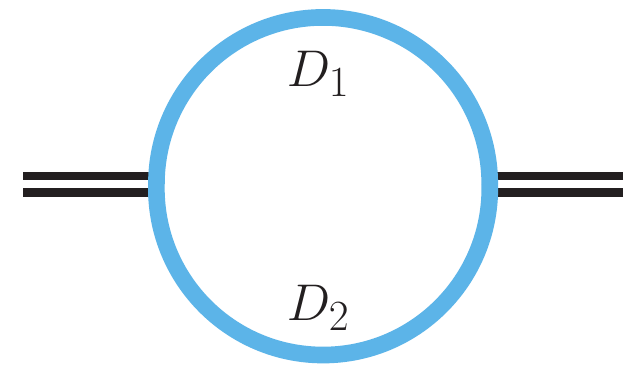}
  \caption{Massive one-loop bubble.}
  \label{fig:bubble1L}
\end{figure}

We describe a simple example for computing a connection matrix for the dual integrals of the one-loop massive bubble family with internal mass $m^2$ depicted in fig.~\ref{fig:bubble1L}.
The family is defined by the proper denominators
\begin{equation}
  D_1 = k^2 - m^2 , \qquad D_2 = (k-p)^2-m^2.
\end{equation}
It has one independent external leg, $p$, that satisfies
\begin{equation}
  p^2 = s.
\end{equation}
The Baikov representation for dual integrals reads
\begin{equation}
  I_{\alpha} = \int \dd^2 \z \, \varphi_{\alpha}(\z) \, u(\z),
\end{equation}
with $\alpha = (\alpha_1,\alpha_2)$ and
\begin{align}
  \varphi_{\alpha}(\z) ={}& \rho_1^{\Theta(\alpha_1 - 1/2)} \, \rho_2^{\Theta(\alpha_2 - 1/2)} \, \frac{1}{z_1^{\alpha_1}\, z_2^{\alpha_2}} \\  u(\z) ={}& \Big(4 m^2 s-s^2+2 s (z_1+z_2)-(z_1-z_2)^2\Big)^{\frac{3-d}{2}}.
\end{align}

The recursive algorithm to calculate intersection numbers requires the variables to be ordered. We choose the ordering where $z_1$ is the innermost layer. We can rewrite the bubble integral family in the iterated form
\begin{align}
   I_{\alpha} ={}& \rho_2^{\Theta(\alpha_2 - 1/2)} \int \dd z_2 \,  \frac{1}{z_2^{\alpha_2}}\, I_{\alpha_1}^{(1)}, \label{eq:bubbletwofoldrec}\\ I_{\alpha_1}^{(1)} ={}& \int \dd z_1 \, u(\z)\, \varphi_{\alpha_1}(z_1). \label{eq:bubbleonefoldinner}
\end{align}
where the integral $I_{\alpha_1}^{(1)}$ has a parametric dependence on $z_2$ and
\begin{equation}
  \varphi_{\alpha_1}(z_1) = \rho_1^{\Theta(\alpha_1 - 1/2)}\, \frac{1}{z_1^{\alpha_1-\rho_1}}.
\end{equation}

We aim to compute the connection matrix $\Omega^{(1)}$ associated to the differential equation satisfied by the $1-$fold integrals $I_{\alpha_1}^{(1)}$:
\begin{equation}
  \frac{\dd}{\dd z_{2}}I^{(1)}_{\alpha_1} = \sum_{\beta \in \textrm{MIs}}\Omega^{(1)}_{\alpha_1 \beta}\, I^{(1)}_{\beta}.\label{eq:Omega1}
\end{equation}
The matrix $\Omega^{(1)}$ is needed to express intersection numbers for the two-fold integrals in eq.~\eqref{eq:bubbletwofoldrec} in terms of intersection numbers for the one-fold integrals in~\eqref{eq:bubbleonefoldinner}.

As a first step, we compute annihilators for the twist containing derivatives with respect to $z_1$ only, since it is the only integration variable for the innermost layer we are considering. We find one first-order annihilator
\begin{equation}
  \hat A = c_0(\z) + c_1(\z)\, \partial_{z_1}
\end{equation}
with the following expression for the polynomial coefficients $c_j(\z)$
\begin{align}
  c_0(\z) ={}& (3-d) (s-z_1+z_2) , \nn c_1(\z) ={}& -4 m^2 s+s^2-2 s (z_1+z_2)+(z_1-z_2)^2.
\end{align}
Integral identities are thus found applying eq.~\eqref{eq:Atmpeq}, which in this case takes the form
\begin{equation}
  \int \dd z_1 u(\z) \Big( c_0(\z)\, \varphi_{\alpha_1} - \partial_{z_1} \big( c_1(\z)\, \varphi_{\alpha_1} \big) \Big) = 0.
\end{equation}
The terms in parentheses are computed by applying the rules in equations~\eqref{eq:dualmon} and~\eqref{eq:dualdiff}.  As an example, for the choice $\alpha_1=1$ the equation reads
\begin{equation}
 \left(-4 m^2 s+s^2-2 s z_2+z_2^2\right) I^{(1)}_{2}-(d-3) (s+z_2) I^{(1)}_{1} = 0,
 \label{eq:dualbubbleI2}
\end{equation}
and can be solved to rewrite for $I^{(1)}_{2}$ in terms of $I^{(1)}_{1}$.  Using this and other identities valid for various integer values of $\alpha_1$, we are able to reduce all integrals $I^{(1)}_{\alpha_1}$ to two master integrals, which can be chosen as
\begin{equation}
  \{I^{(1)}_{1}, I^{(1)}_{0}\}.
\end{equation}

To obtain the connection matrix $\Omega^{(1)}$, we derive the differential operator $\hat O_{z_2}$, which realizes the differentiation with respect to $z_2$ for the twist $u(\z)$ as in eq.~\eqref{eq:Ode}.  Once again, since we are considering the innermost integration in $z_1$, the operator must be polynomial in $z_1$ and can only contain derivatives with respect to $z_1$. On the other hand, it may have a rational dependence on $z_2$, $m^2$ and $d$. We find the first-order operator
\begin{equation}
  \hat O_{z_2} = c_0^{(z_2)}(\z) + c_1^{(z_2)}(\z)\, \partial_{z_1}
\end{equation}
with
\begin{align}
  c_0^{(z_2)}(\z) ={}& -\frac{d-3}{2 (m^2+z_2)} \nn
  c_1^{(z_2)}(\z) ={}& -\frac{2 m^2-s+z_1+z_2}{2 (m^2+z_2)}.
\end{align}
In particular, applying it to the master integrals $I^{(1)}_{0}$ and $I^{(1)}_{1}$, using~\eqref{eq:dIalpha} and again equations~\eqref{eq:dualmon} and~\eqref{eq:dualdiff}, we obtain
\begin{align}
 &\partial_{z_2} I^{(1)}_{0} = \frac{(4-d) I^{(1)}_{0}}{2 (m^2+z_2)}+\frac{(2 m^2-s+z_2) I^{(1)}_{1}}{2 (m^2+z_2)}, \nn
 &\partial_{z_2} I^{(1)}_{1} = \frac{(3-d) I^{(1)}_{1}}{2 (m^2+z_2)}+\frac{(-2 m^2+s-z_2) I^{(1)}_{2}}{2 (m^2+z_2)}.
\end{align}
By inserting the reduction of $I^{(1)}_{2}$ to master integrals, found by solving~\eqref{eq:dualbubbleI2}, we finally obtain the system of first order differential equation in eq.~\eqref{eq:Omega1} where the connection matrix reads
\begin{equation}
  \Omega^{(1)} = \begin{pmatrix}
    \frac{(d-3) (s-z_2)}{(s-z_2)^2-4 m^2 s} & 0 \\
    \frac{2 m^2-s+z_2}{2 (m^2+z_2)} & -\frac{d-4}{2 (m^2+z_2)}
  \end{pmatrix}.
\end{equation}

\subsubsection*{Additional examples}

\begin{figure}
     \centering
     \begin{subfigure}[b]{0.23\textwidth}
         \centering
         \includegraphics[width=0.6\textwidth]{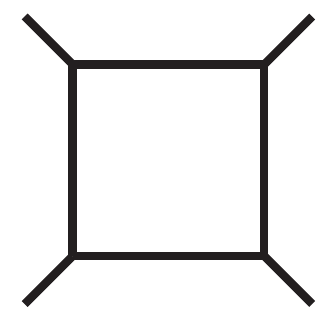}
         \caption{One-loop box}
         \label{fig:ints-box}
     \end{subfigure}\hspace{-0.1cm}%
     \begin{subfigure}[b]{0.23\textwidth}
         \centering
         \includegraphics[width=0.6\textwidth]{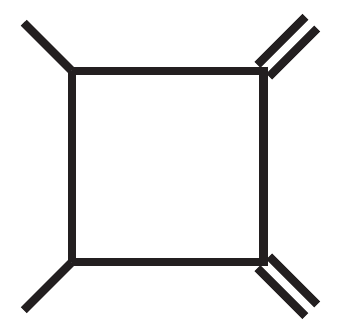}
         \caption{One-loop hard box}
         \label{fig:ints-hardbox}
     \end{subfigure}\hspace{-0.1cm}%
     \begin{subfigure}[b]{0.25\textwidth}
         \centering
         \includegraphics[width=\textwidth]{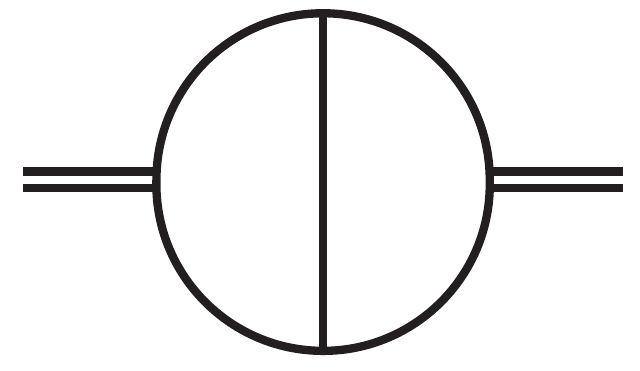}
         \caption{Kite}
         \label{fig:ints-kite}
     \end{subfigure}\hspace{0.3cm}%
     \begin{subfigure}[b]{0.25\textwidth}
         \centering
         \includegraphics[width=\textwidth]{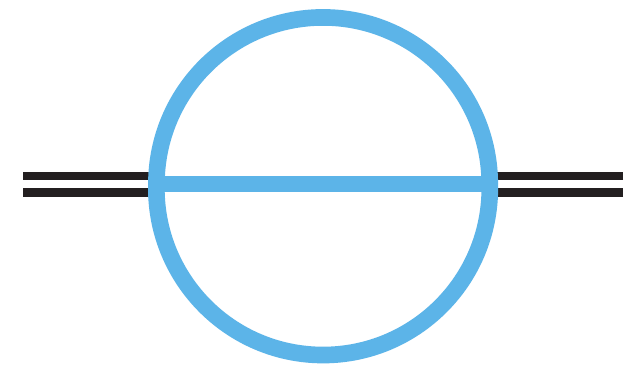}
         \caption{Equal-mass sunrise}
         \label{fig:ints-sunrise}
     \end{subfigure}
        \caption{Integral families with regulated denominators considered in ref.~\cite{Fontana:2023amt}, for which connection matrices of inner integration layers have been checked against the method proposed in this paper.}
        \label{fig:three graphs}
\end{figure}

We tested this approach on all the connection matrices for all the integral families with non-trivial (regulated) denominators appearing in the applications of ref.~\cite{Fontana:2023amt}.  The diagrams corresponding to their top sectors are depicted in fig.~\ref{fig:three graphs}.  These tests are published with the \textsc{CALICO} program.

\section{Loop integrals as twisted Mellin moments}
\label{sec:twisted}

In this paper, we define \emph{twisted Mellin moments} to be families of integrals of the form in eq.~\eqref{eq:Ialpha} with integration contour $z_i\in (0,\infty)$ and
\begin{equation}
  \label{eq:mellin}
  \varphi_{\alpha} = \Big(\prod_{j=1}^n\frac{1}{\Gamma(\alpha_j)}\Big)\, \z^{\alpha-1},
\end{equation}
where $\alpha-1 = (\alpha_1-1,\ldots,\alpha_n-1)$.    Several parametric representations of loop integrals (see eq.~\eqref{eq:loopint}) have this form, where we identify
\begin{equation}
  J_\alpha \propto I_\alpha, \label{eq:mellinproptoloop}
\end{equation}
where the proportionality factor depends on the specific representation and on the exponents $\alpha$.

We note that, while the space of functions $\varphi_{\alpha}$ is closed under monomial multiplication and differentiation, hence satisfying our assumptions, the $\Gamma(\alpha_j)$ factors need to be adjusted accordingly when converting monomials into functions $\varphi_{\alpha}$.  In particular, a derivative of the twist can be converted into
\begin{equation}
  \label{eq:mellinibp}
  \int_0^\infty \dd^n \z\, \varphi_{\alpha}(\z)\, \partial_{z_j} u(\z) = - \int_0^\infty \dd^n \z\, \varphi_{\alpha-\hat e_j}(\z)\, u(\z),
\end{equation}
where, after an IBP, we used $\Gamma(\alpha_j)=(\alpha_j-1)\, \Gamma(\alpha_j-1)$.
It would naively appear that, in the IBP we just used, we neglected a boundary term at $z_j=0$ in the special case $\alpha_j=1$.  However, one can check that the equality one finds assuming $\alpha_j>1$ (such that the boundary term does not contribute) is analytic in $\alpha$ and can be continued\footnote{For the special case $\alpha_j=1$, one can easily check that only the boundary term at $z_j=0$ contributes to the IBP identity.  However, in parametric representations of Feynman integrals, setting $\alpha_j=1$ and $z_j=0$ in the whole integrand (including the twist) corresponds to removing the $j$-th denominator, which in turn corresponds to setting the exponent $\alpha_j=0$ in eq.~\eqref{eq:loopint}.  Hence, eq.~\eqref{eq:mellinibp} is still valid in the limit $\alpha_j\to 1$, once written in terms of $J_\alpha$.} to $\alpha_j=1$ and, in fact, also to $\alpha_j\leq 0$.  Indeed, the dimensionally regulated integrals $J_\alpha$ can be analytically continued in the exponents $\alpha_j$ and, for generic $d$, have non-singular limits for integer values of $\alpha_j$.  This is also true for the integrals $I_\alpha$ we just defined, since the proportionality factor in eq.~\eqref{eq:mellinproptoloop} is non-singular in these limits, as we will see below.  We can thus write template equations from annihilators for generic exponents, neglecting boundary terms in IBPs, and then analytically continue them to any value of $\alpha_j$, including zero or negative integers.

\subsection{Lee-Pomeransky representation}
The Lee-Pomeransky representation of loop integrals is related to the twisted Mellin moments with the special choice of twist
\begin{equation}
  \label{eq:LP}
  u(\z) = G(\z)^{-d/2}
\end{equation}
where $d$ is a regulator.  The loop integrals in eq.~\eqref{eq:loopint} can be written in this form, up to a prefactor
\begin{equation}
  J_\alpha = (-1)^{|\alpha|}\frac{\Gamma\left(\frac{d}{2}\right)}{\Gamma\left(\frac{(\ell+1) d}{2}-|\alpha|\right)}\, I_\alpha,
\end{equation}
by identifying $d$ with the space-time dimension and the polynomial $G(\z)$ as the sum of the two Symanzik polynomials $\U$ and $\F$ (see Appendix~\ref{sec:feynLPpar} for a definition),
\begin{equation}
  G(\z) = \U(\z) + \F(\z).
\end{equation}
This is known as the Lee-Pomeransky parametrization of loop integrals~\cite{Lee:2013hzt}. Parametric annihilators in this integral representation have already been extensively discussed in~\cite{Bitoun:2017nre}.

In this section, as well as in the \textsc{CALICO} program presented in section~\ref{sec:calico}, we deal with Mellin moments of the form in eq.~\eqref{eq:mellin}.  Converting a reduction of the form in eq.~\eqref{eq:red} for Mellin integrals into one for the corresponding loop integrals
\begin{equation}
  J_\alpha = \sum_{\beta\in \textrm{MIs}} \tilde c_{\alpha\beta}\, J_\beta
\end{equation}
is straightforward by identifying
\begin{equation}
  \tilde c_{\alpha\beta} = (-1)^{|\beta|-|\alpha|}\, \frac{\Gamma\left(\frac{(\ell+1) d}{2}-|\beta|\right)}{\Gamma\left(\frac{(\ell+1) d}{2}-|\alpha|\right)}\, c_{\alpha\beta}.
\end{equation}
Note that, because both $|\alpha|$ and $|\beta|$ are integers, the arguments of the two Gamma functions in the ratio above differ by an integer number, hence it is always a rational function of $d$.  The latter can be found by recursively applying the relation $\Gamma(t+1)=t\, \Gamma(t)$ to shift the argument of one of the two $\Gamma$ functions until they simplify yielding simple rational factors. In practice, we find more convenient to apply this relation directly at the level of the template equations, to translate the identities in terms of loop integrals before they are solved.

The Lee-Pomeransky parametrization of loop integrals has a number of advantages compared to other representations, such as Baikov's or momentum space.  First, it only involves $n$ integration variables, where $n$ is the number of loop denominators, with no need of introducing ISPs --- although the latter can also be added if required.  Moreover, the degree of the polynomial $G$ is always $\deg G = \ell+1$ at $\ell$ loops (see e.g.\ Appendix~\ref{sec:feynLPpar}).  For comparison, the degree of the Baikov polynomial is $\deg B = \min(\ell+e,2\ell)$, where $e$ is the number of independent external momenta, hence $\deg B\geq \deg G$ for $e\geq 1$.  This makes this method particularly efficient in some contexts, as already observed in the literature (see e.g.~\cite{vonManteuffel:2019wbj}).  Its main drawback is that, in the presence of ISPs, the number of independent annihilators is comparatively large and it is thus less convenient than other representations.

\subsection{Schwinger representation}
The Schwinger representation of loop integrals is related to twisted Mellin moments with twist
\begin{equation}
  \label{eq:schwinger}
  u(\z) = \exp \big[ -\F(\z)/\U(\z)\big]\, \U(\z)^{-d/2}
\end{equation}
and the integral $I_\alpha$ corresponds to the loop integral in eq.~\eqref{eq:loopint} up to a sign
\begin{equation}
  J_\alpha = (-1)^{|\alpha|}\, I_\alpha.
\end{equation}
As before, $d$ is the space-time dimension, while $\U$ and $\F$ are the Symanzik polynomials.

The Schwinger representation shares many properties with the Lee-Pomeransky representation, including the fact that it consistently works without the need of ISPs --- although, in both cases, they can be included when needed. Moreover, we empirically find that annihilators generally have lower degree in this representation, as compared to Lee-Pomeransky.  When a full set of ISPs is used, for instance, first-order annihilators in the Schwinger representation all have degree one. However, due to a more complex twist, which also includes an exponential factor, finding such annihilators can sometimes be harder, depending on the specific problem. Moreover, we empirically find that we need to use second-order annihilators in this representation more often than in others, which can significantly add complexity to a calculation. On the other hand, template identities are generally fewer and simpler in this representation, compared to Lee-Pomeransky or others, hence they often yield simpler systems of equations and more efficient reductions to master integrals.

\subsection{Examples}

The methods described in this work have been checked in several non-trivial examples using both the Schwinger and the Lee-Pomeransky parametrization. We tested reductions both with and without including a full set of ISPs in the definition of the integral families. The relations we derive agree with those obtained with the Laporta method in momentum space.

In particular, by using annihilators obtained by defining, for each sector, a family of twisted Mellin moments with only its proper denominators and no ISP, one can efficiently obtain reductions with very high powers of denominators~\cite{vonManteuffel:2019wbj}.  These can be, in turn, useful because some of these integrals have good properties such as (quasi-)finiteness~\cite{Panzer:2014gra,vonManteuffel:2014qoa} or uniform trascendental weight~\cite{Henn:2013pwa}.

Several non-trivial examples have been checked via a preliminary interface between the \textsc{CALICO} package (presented in this work) and the private \textsc{FFIntRed} package for integral reduction, which we plan to share in a future work.
\begin{figure}
  \centering
  \includegraphics[width=6cm]{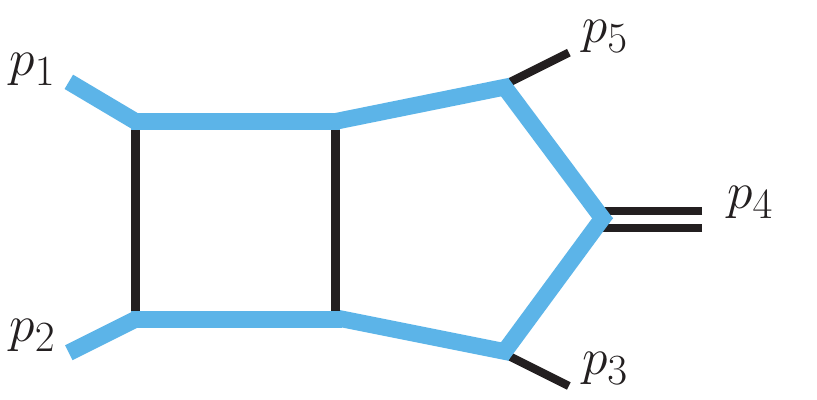}
  \caption{Pentabox contributing to top-pair plus Higgs production.}
  \label{fig:ttH}
\end{figure}
One of them is the reduction of all integrals with trivial numerators (i.e.\ $s=1$, defined in section~\ref{sec:loop}) and up to 3 dots ($u=3$) for the integral family in fig.~\ref{fig:ttH}. The latter contributes to top-pair plus Higgs production at hadron colliders at two loops accuracy.  We computed the annihilators numerically and used them for generating a linear system solving the integral relations, combined with symmetry relations found with the \textsc{FFIntRed} package.  We tested this, on a modern laptop, using both the Lee-Pomeransky and the Schwinger parametrization.  For the latter, computing the annihilators and figuring out the correct seeding --- which turned out had to include seeds with $s=1$ --- was more complicated, also due to the need of using some second-order annihilators.  However, after the system had been generated and independent equations had been filtered out by the \textsc{FiniteFlow} solver, this method combined with the Schwinger parametrization yielded an extremely efficient system of equations. This contained about 20'000 equations and could be solved numerically in a few hundredths of a second on a laptop, which is orders of magnitude more efficient than what we were able to obtain using other approaches or integral representations.

While we leave a detailed description of complex examples of this kind to future works, the core ideas behind this approach are also illustrated in the two simple applications we discuss in detail below. An implementation of them can be found in the public repository of the \textsc{CALICO} program.

\subsubsection*{The two-loop sunrise with different masses}
\begin{figure}
  \centering
  \includegraphics[width=5.2cm]{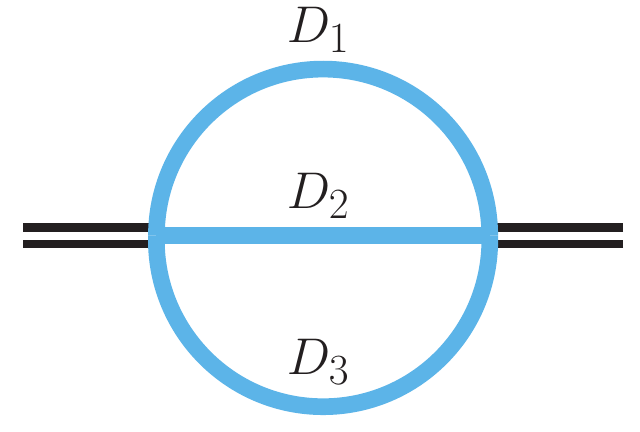}
  \caption{Two-loop sunrise with different internal masses.}
  \label{fig:sunrise}
\end{figure}
As a first simple example, consider the two-loop sunrise with different masses (depicted in fig.~\ref{fig:sunrise}), i.e.\ the family of loop integrals defined by the denominators
\begin{align}
  D_1 ={}& k_1^2-m_1^2 \nn
  D_2 ={}& k_2^2-m_2^2 \nn
  D_3 = {}& (k_1+k_2 - p)^2 - m_3^2.\label{eq:sunrisedens}
\end{align}
These integrals depend on four invariants, namely the three squared masses $m_j^2$ and
\begin{equation}
  s \equiv p^2. \label{eq:sdefp2}
\end{equation}
We immediately observe that, in order to produce integral identities via IBPs in momentum space for this family, two ISPs are needed.  However, using annihilators in the Lee-Pomeranski or Schwinger parametrization of loop integrals, ISPs are not needed.

Using the Lee-Pomeranski representation, we find 3 generators for first-order annihilators of degree 2 and 6 of degree 3.  Using the Schwinger representation, we find instead 5 first-order generators, as well as one second-order generator, all of which have degree 2. This second-order annihilator is needed to obtain a complete reduction to master integrals in the Schwinger representation, which would yield one additional master integral otherwise. As explained, each generator yields a template equation for the integral family, using eq.~\eqref{eq:Atmpeq}.

We further observe that this family has one top sector and 3 non-zero subsectors. The proper denominators of these subsectors factorize into products of one-loop vacuum integrals (a.k.a.\ tadpole integrals). In principle, we can limit ourselves to using the same template identities defined above, considering seeds with zero or negative exponents as well. However, for the subsectors, we also consider identities obtained from annihilators of one-loop vacuum integrals identified by the denominator
\begin{align}
  D_0=k^2-m^2\label{eq:tadden}
\end{align}
with $m$ replaced by any of the $m_j$.  For this, we obtain one first-order generator, of degree 2 for Lee-Pomeranski and degree 1 for Schwinger.  After replacing $m\to m_j$ with $j=1,2,3$ we thus obtain 3 more template identities valid for subsectors.  Of course, this is of no importance in this simple example, but the same principle applied to more complex examples may substantially decrease the complexity of a reduction.

For all sectors we use seed integrals with trivial numerators and find 7 MIs in the whole family, which can be chosen as
\begin{equation}
\{J_{112}, J_{121},
 J_{211}, J_{111},
 J_{011}, J_{101},
 J_{110}\}.
\end{equation}
Differential operators $\hat O_s$ and $\hat O_{m_j^2}$ are found following section~\ref{sec:de}, which we use to find DEs satisfied by the MIs.  These have been checked against a calculation that uses the Laporta algorithm.

\subsubsection*{Equal-mass $\ell$-loop banana integrals}
\begin{figure}
  \centering
  \includegraphics[width=5.2cm]{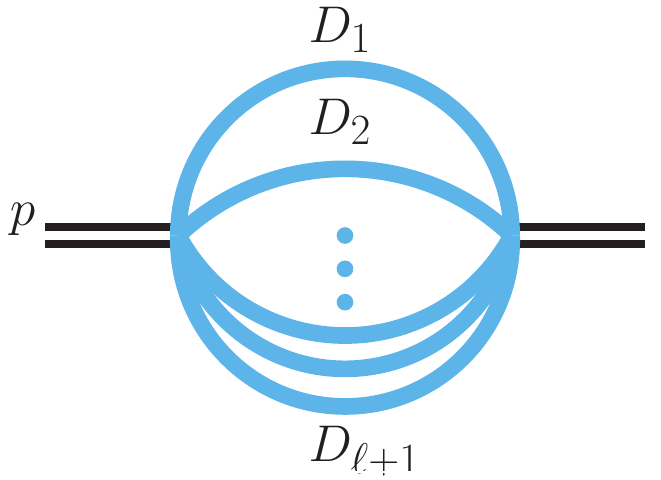}
  \caption{$\ell$-loop banana graph.}
\end{figure}
The $\ell$-loop banana integral family with internal mass $m$ is defined by the set of denominators
\begin{align}
  D_j ={}& k_j^2-m^2 \quad \textrm{for }j=1,\ldots,\ell \nn
  D_{\ell+1} = {}& (k_1+\cdots + k_\ell - p)^2 - m^2.\label{eq:bananadens}
\end{align}
This family depends on two invariants, namely $m^2$ and $s$, with the latter defined as in eq.~\eqref{eq:sdefp2}.
These integrals are of great interest due to their analytic structure. For $\ell \geq 2$, they are indeed among the simplest non-factorizable $\ell$-loop integrals which cannot be expressed in terms of generalized polylogarithms.

Despite their apparent algebraic simplicity, deriving differential equations for these integrals can actually become challenging as the number of loops $\ell$ grows, when using IBPs in momentum space. This is due to the need of adding a large number of ISPs (namely $(\ell+2)(\ell-1)/2$), which grows quadratically in $\ell$, to have a complete set of generalized denominators. As an example, at six loops we would need 20 ISPs. The representations of loop integrals reviewed in this section, however, do not require introducing ISPs and can work just the $\ell+1$ denominators in eq.~\eqref{eq:bananadens}, drastically simplifying the problem. We tested this examples for several choices of the loop order $\ell$, finding the general patterns which we describe below.  We recall that alternative efficient algorithms for computing the Picard-Fuchs operator of $\ell$-loop banana graphs exist (see e.g.~\cite{Vanhove:2014wqa,Bonisch:2021yfw,Pogel:2022vat}) but the one we use here is much more general and not limited to this integral family.

In the Lee-Pomeransky representation we find first-order annihilators up to degree $\ell+1$.  In the Schwinger parametrization we find first-order annihilators up to degree $\ell$, as well as one second-order annihilator of degree 2 which is required for a complete reduction.  Besides the top sector, we have $\ell+1$ subsectors, whose integrals with trivial (i.e.\ constant) numerators factorize as products of $\ell$ one-loop integrals with one denominator as in eq.~\eqref{eq:tadden}.

This integral family also has a clear symmetry.  Integrals $\I_\alpha$ are symmetric under any permutation of the indexes $\alpha_1,\ldots,\alpha_n$ (where here $n=\ell+1$). We can simply implement this by sorting the indexes identifying an integral, say, from higher to lower.  By doing this, all the integrals are mapped to the two sectors $S_{1\cdots 1}$ and $S_{1\cdots 1 0}$.

By combining  these symmetries with identities generated using annihilators and seeds with trivial numerator, we find $\ell+1$ master integrals, that can be chosen as
\begin{equation}
  \{\I_{2\cdots 2 1 1}, \I_{2\cdots 2 1 1 1}, \ldots, \I_{2 1\cdots 1}, \I_{1\cdots 1}, \I_{1\cdots 1 0}\}.
\end{equation}

Differential operators $\hat O_s$ and $\hat O_{m^2}$ are also found as first-order operators with maximum degree 2 (or 1 for $\ell=1$).  Combined with the identities above, we thus easily find DEs satisfied by the MIs.  We successfully checked these against a calculation that uses the Laporta algorithm in momentum space up to $\ell=5$ loops, although the approach described here is orders of magnitude more efficient, since it does not need ISPs.  Indeed, even at $\ell=6$ loops we are able to find annihilators and template identities in a couple of minutes on a modern laptop, after which deriving the DEs (including the generation of the system of equations, the derivative operators and the analytic reconstruction of the matrices) takes no more than a few seconds.

Although this is a relatively simple example, it shows that for specific applications the approach of annihilators can outperform state-of-art alternatives, as already seen in the literature (see e.g.~\cite{vonManteuffel:2019wbj}).

\section{The \textsc{CALICO} Mathematica package}
\label{sec:calico}
In this section we give a general description of the \textsc{CALICO}
program and its features.  The package is open source and it is available at
\begin{equation*}
  \textrm{\url{https://github.com/fontana-g/calico}}
\end{equation*}
where several examples with detailed comments can also be found.  Here we describe the core features of the package and we refer to the builtin examples, and comments therein, for more details about using it in non-trivial applications. These include most of the examples illustrated in the previous sections.

The package depends on the \textsc{FiniteFlow}~\cite{Peraro:2019svx} program,\footnote{\textsc{CALICO}'s performance can benefit from new features of an upcoming release of~\textsc{FiniteFlow} which, at the time of writing, are available in the experimental branch \texttt{exp} of the public \textsc{FiniteFlow} repository.} which is used for solving linear identities constraining the coefficients of annihilators and other differential operators (see also section~\ref{sec:ffdetails}).

We load the package with
\begin{center}
\begin{lstlisting}
  <<CALICO%\textasciigrave%
\end{lstlisting}
\end{center}
All public symbols exported by the package are prefixed by \texttt{CAT} (as in ``Computing AnnihilaTors'') to avoid conflicts with other packages or builtin symbols.

The most important function of the package is arguably \texttt{CATAnnihilator}, which is used as follows
\begin{center}
\begin{lstlisting}
  ann = CATAnnihilator[$u(\z)$,{$z_1$,$\ldots$,$z_n$},$d_{\textrm{max}}$,$o_{\textrm{max}}$];
\end{lstlisting}
\end{center}
where $u(\z)$ is the twist, while $d_{\textrm{max}}$ and $o_{\textrm{max}}$ are the maximum degree and order of the generators of the annihilators to be found.  The output of the function contains the coefficients $c_{j_1 j_2\cdots}(\z)$ which define the annihilators as in eq.~\eqref{eq:Ao}. Its exact form is reported later, but it is unlikely to be important for most users, since we already provide functions which use the output \texttt{ann} to produce integral identities for many types of integral families.  However, it is worth mentioning that
\begin{center}
\begin{lstlisting}
  Length[ann[[$o$,$d+1$]]]
\end{lstlisting}
\end{center}
returns the number of generators of order $o$ and degree $d$ which have been found.  The annihilators returned by this function are, by default, cast as polynomials in both $\z$ and in the free parameters appearing in the twist. This yields integral identities that are easier to manipulate, but in principle annihilators only need to be polynomials in $\z$. Casting them in the default form however requires additional algebraic manipulations, such as computing the least common multiple of polynomials in the free parameters. The option \texttt{"PolynomialInParameters"->False} can be specified, saving some algebraic manipulations, if annihilators are not needed to be cast in this form. We also clarify that \textsc{CALICO} does not provide a way of choosing $d_{\textrm{max}}$ and $o_{\textrm{max}}$. In most applications, one may wish to adjust these values either until no more annihilators are found beyond a certain $o$ and $d$ (see also the previous code block), or until one can check via other means that the generators that have been found are sufficient to solve a certain problem.

Sometimes, one may need to compute annihilators in multiple steps.  As an example, after computing \texttt{ann} as before, we may wish to increase $d_{\textrm{max}}$ or $o_{\textrm{max}}$ to seek additional solutions, without recomputing again the generators that are already in \texttt{ann}.  This can be achieved using
\begin{center}
\begin{lstlisting}
  annextra = CATAnnihilator[$u(\z)$,{$z_1$,$\ldots$,$z_n$},$d_{\textrm{max}}'$,$o_{\textrm{max}}'$,
                            $\texttt{"KnownSolutions"}$->ann];
\end{lstlisting}
\end{center}
after which \texttt{annextra} will contain generators up to degree $d_{\textrm{max}}'$ and order $o_{\textrm{max}}'$ that are independent of those in \texttt{ann}. A similar call can also be used to exclude solutions generated by operators \texttt{ann} that are known or defined by any other mean.  The options \texttt{"MinDegree"->}$d_{\textrm{min}}$ and \texttt{"MinOrder"->}$o_{\textrm{min}}$ are available to specify the minimum degree and order at which the search for a solution is started (by default they are 0 and 1 respectively).  The two solutions can thus be merged using
\begin{center}
\begin{lstlisting}
  ann = CATAnnihilatorMerge[ann,annextra];
\end{lstlisting}
\end{center}
after which \texttt{ann} will include generators previously contained either \texttt{ann} or \texttt{annextra}.

The other main function of the package is \texttt{CATDiffOperator}, which is used to compute differential operators in the integration variables that are equivalent to differentiation of the twist with respect to an external free parameter.  If $x_1,\ldots,x_m$ are external parameters, then
\begin{center}
\begin{lstlisting}
  diffop = CATDiffOperator[$u(\z)$,{$z_1,\ldots,z_n$},{$x_1,\ldots,x_m$},$d_{\textrm{max}}$,$o_{\textrm{max}}$,
                           $\texttt{"KnownAnnihilatorSolutions"}$->ann];
\end{lstlisting}
\end{center}
returns a list of $m$ elements having, at position $i$, analytic data about the differential operator $\hat O_{x_i}$ defined as in eq.~\eqref{eq:Ode}, or \texttt{CATImpossible} if no such operator was found  (in such case one may wish to try again with higher values of $d_{\textrm{max}}$ or $o_{\textrm{max}}$).  The inputs other than $x_i$ have the same meaning as before. The option \texttt{"KnownAnnihilatorSolutions"} in the second line is not required but it can improve performance, since \textsc{CALICO} will use known annihilators \texttt{ann} of the twist $u(\z)$ to trim the ansatz for the differential operators.

Both functions we just described use the functional reconstruction techniques of \textsc{FiniteFlow} to reconstruct analytic results from numerical evaluations over finite fields.  The user can specify the maximum total degree (in the free parameters, including the regulators) and the maximum number of prime fields to use for the reconstruction via the options \texttt{"MaxRecDegree"} and \texttt{"MaxRecPrimes"} respectively. Similarly, the option \texttt{"NThreads"} can be specified to control the number of threads used in multivariate reconstructions. The option \texttt{"Substitutions"} takes instead a list of substitutions in the free parameters that is performed right before solving the syzygy or polynomial decomposition equations. It may be used e.g.\ to set one dimensional variable to one (later recovering its dependence via dimensional analysis), which improves performance of reconstruction algorithms, or to test the calculation for rational numerical values of the free parameters.

The calculation of annihilators and differential operators might take a substantial amount of time for very complex applications. \textsc{CALICO} can optionally print information about the ongoing calculation.  This is done when the verbosity is set to \texttt{True}, which is achieved calling
\begin{center}
\begin{lstlisting}
  CATVerbose[True];
\end{lstlisting}
\end{center}
before the call to these functions.

The twist $u(\z)$ taken as input by the functions above can be passed just by using its \emph{analytic expression}, but \textsc{CALICO} has many functions for building twists for a wide variety of integral families.  These cast the twist into a form that is optimized for the analytic manipulations performed internally by these functions, as briefly mentioned in section~\ref{sec:ffdetails}.  The wrapper function \texttt{CATTwist} is used to signal this optimized form to \textsc{CALICO}.  Users can also build twists in such a form, as described later when more advanced use cases are discussed.

We will now give an overview of basic functions that can be used to
\begin{itemize}
\item define the twist $u(\z)$ and the polynomials it depends on;
\item convert annihilators into template equations, for various forms of $\varphi_\alpha$;
\item generate lists of seeds, to which the template equations can be applied;
\item convert differential operators into derivatives of integrals, as functions of their exponents, for various forms of $\varphi_\alpha$.
\end{itemize}

We first review the functions we provide for building polynomials and twists related to various integral families.

For the Baikov parametrization
\begin{center}
\begin{lstlisting}
  {baikov,zs} = CATBaikovPoly[
    {{$l_1$,$m_1^2$},$\ldots$,{$l_n$,$m_n^2$}},  (* generalized denominators *)
    {$k_1$,$\ldots$,$k_\ell$}, (* loop momenta *)
    {$p_1$,$\ldots$,$p_e$},    (* linearly independent external momenta *)
    replacements,
    z
  ]
\end{lstlisting}
\end{center}
sets \texttt{baikov} to the Baikov polynomial and \texttt{zs} to the list of integration variables
\begin{equation*}
  \texttt{\{z[1],$\ldots$,z[$n$]\}}.
\end{equation*}
The pairs $\{l_j,m_j^2\}$ identify the generalized denominators in momentum space, as in eq.~\eqref{eq:loopden}, with $l_j$ being linear combinations of the $\ell$ loop momenta $k_i$ and the $e$ independent external momenta $p_i$ (excluding the one fixed by momentum conservation which must not appear in the $l_j$).  Bilinear generalized denominators, as in eq.~\eqref{eq:loopdenbl}, may be specified using $\{l_j,v_j,m_j^2\}$ instead.  The input \texttt{replacements} should be a list of rules of the form $p_i p_j \rightarrow \ldots$ converting scalar products (specified as normal products in the rules) of external momenta into invariants.  The corresponding twist is obtained using\footnote{Note that \texttt{CATBaikov} only depends on the Baikov polynomial, hence it does not include the external prefactor, which depends on the kinematic invariants and is needed to derive the differential operator $\hat O_x$.  For this purpose, one should use \texttt{CATMultiBaikov} instead, including all kinematic-dependent factors.}
\begin{center}
\begin{lstlisting}
  CATBaikov[baikov,$d$,$\ell$,$e$,zs]
\end{lstlisting}
\end{center}
with $d$ being the number of space-time dimensions (typically a symbol $d$ or $4-2\epsilon$), $\ell$ the number of loops and $e$ the number of independent external momenta.  Similarly, the twist for dual loop integrals in the Baikov representation is obtained with
\begin{center}
\begin{lstlisting}
  CATBaikovDual[baikov,$d$,$\ell$,$e$,zs]
\end{lstlisting}
\end{center}
whose inputs have the same meaning.

For the integral parametrizations discussed in section~\ref{sec:twisted}, we need the Symanzik polynomials $\U$ and $\F$, as well as $G=\U+\F$.  In \textsc{CALICO} we compute them using
\begin{center}
\begin{lstlisting}
  {u,f,g,zs} = CATUFGPolys[
    {{$l_1$,$m_1^2$},$\ldots$,{$l_n$,$m_n^2$}},
    {$k_1$,$\ldots$,$k_\ell$},
    replacements,
    z
  ];
\end{lstlisting}
\end{center}
where the inputs have the same meaning as above.  This sets the symbols \texttt{u}, \texttt{f} and \texttt{g} to $\U(\z)$, $\F(\z)$ and $G(\z)$ respectively and assigns the list of integration variables to \texttt{zs}, as before. The twist is returned by the call
\begin{center}
\begin{lstlisting}
  CATLP[g,$d$,zs]
\end{lstlisting}
\end{center}
for the Lee-Pomeransky representation and
\begin{center}
\begin{lstlisting}
  CATSchwinger[u,f,$d$,zs]
\end{lstlisting}
\end{center}
for the Schwinger representation.

Another function, that can define the twist in a wide variety of cases, is
\begin{center}
\begin{lstlisting}
  CATMultiBaikov[{$B_1(\z)$,$B_2(\z)$,$\ldots$},{$\gamma_1$,$\gamma_2$,$\ldots$},$\z$];
\end{lstlisting}
\end{center}
to define a twist with the form in eq.~\eqref{eq:u} or
\begin{center}
\begin{lstlisting}
  CATMultiBaikov[{$B_1(\z)$,$B_2(\z)$,$\ldots$},{$\gamma_1$,$\gamma_2$,$\ldots$},$F(\z)$,$\z$]
\end{lstlisting}
\end{center}
to define a twist with the form in eq.~\eqref{eq:uexp} if $F(\z)$ is a polynomial.  This form covers, in principle, all cases illustrated in this paper except for Schwinger parametrization, although we generally prefer dedicated routines for most of them.  \texttt{CATMultiBaikov} is however the function we use for hypergeometric functions and the loop-by-loop Baikov parametrization.  For the latter, \textsc{CALICO} does not provide functions for computing the polynomials $B_j(\z)$ and the exponents $\gamma_j$ but this functionality is already available in the public \textsc{BaikovPackage}~\cite{Frellesvig:2024ymq}.  In the builtin example that uses this parametrization, we explicitly show how the input for \textsc{CALICO} can be generated using \textsc{BaikovPackage}.

Once the twist has been computed, it can be used to compute parametric annihilators and differential operators $\hat O_x$ as already discussed above.  We thus typically need to turn these into integral identities.  This translation effectively implements equations~\eqref{eq:Atmpeq} and~\eqref{eq:dIalpha}, which depend on the specific form of the integrand $\varphi_\alpha$.  \textsc{CALICO} has several functions which perform this translation for different forms of $\varphi_\alpha$.  For this purpose, the package uses the symbolic function
\begin{center}
\begin{lstlisting}
  CATInt[$F$,{$\alpha_1$,$\ldots$,$\alpha_n$}]
\end{lstlisting}
\end{center}
to represent either $\varphi_\alpha$ or the corresponding integral $I_\alpha$, identified by the multi-index of exponents $\alpha = (\alpha_1,\ldots,\alpha_n)$.  The first argument $F$ can be any symbol or expression (even a string) and it is used to identify the integral family, so that integrals belonging to more than one family can co-exist in an expression.  The function \texttt{CATInt} also automatically performs the substitutions\footnote{After these substitutions are performed, prefactors might need to be manually adjusted, depending on the explicit form of $\varphi_\alpha$ (see e.g.\ eq.~\eqref{eq:phidual} and~\eqref{eq:mellin}).  This is only relevant when users implement their own identities, since \textsc{CALICO} already takes care of adjusting prefactors in the template identities it generates.}
\begin{align*}
  & \texttt{CATInt[$F$,$\alpha$] CATInt[$F$,$\beta$]} \rightarrow \texttt{CATInt[$F$,$\alpha+\beta$]} \nn
  & \texttt{CATInt[$F$,$\alpha$]$^n$} \rightarrow \texttt{CATInt[$F$,$n \alpha$]}
\end{align*}
that are consistent with $\alpha_j$ being exponents of rational factors of the integrands $\varphi_\alpha$.  In the following, we describe \textsc{CALICO}'s functions that convert annihilators \texttt{ann} and differential operators $\hat O_x$, denoted as \texttt{diffop} as before, into integral identities of a family $F$.  Moreover, $\z$ corresponds to the list of integration variables, while $\x=\{x_1,\ldots,x_m\}$ is a list of free parameters for which we need differential operators $\hat O_{x_j}$.

The simplest case is when $\varphi_\alpha$ is just a (Laurent) monomial
\begin{equation}
  \varphi_\alpha(\z) = \z^\alpha.
\end{equation}
In such case, template integral identities can be obtained from annihilators using
\begin{center}
\begin{lstlisting}
  tmpids = CATMonIdsFromAnnihilators[$F$,ann,$\z$];
\end{lstlisting}
\end{center}
The returned value \texttt{tmpids} is an anonymous function~\cite{MathFunctionDoc}
of the exponents such that, for each seed $\alpha$
\begin{center}
\begin{lstlisting}
  tmpids[$\alpha_1$,$\ldots$,$\alpha_n$]
\end{lstlisting}
\end{center}
returns a list of linear combinations of integrals which vanish, namely the l.h.s.\ of eq.~\eqref{eq:Atmpeq} for each generator.  Similarly, derivatives of integrals can be generated with
\begin{center}
\begin{lstlisting}
  deriv = CATMonIdsFromDiffOperators[$F$,diffop,$\z$];
\end{lstlisting}
\end{center}
whose return value \texttt{deriv} is also an anonymous function of the exponents. For each seed $\alpha$
\begin{equation*}
  \texttt{deriv[$\alpha_1$,$\ldots$,$\alpha_n$]} = \{\partial_{x_1}\I_\alpha,\ldots,\partial_{x_m}\I_\alpha\},
\end{equation*}
with each entry obtained as the r.h.s.\ of eq.~\eqref{eq:dIalpha}.

Completely analogous functions are available for different forms of $\varphi_\alpha$.  Since they have the same behaviour and take the same inputs as the ones we just described, we simply list them here.
\begin{itemize}
\item For inverses of (Laurent) monomials
  \begin{equation}
    \varphi_\alpha(\z) = \z^{-\alpha},
  \end{equation}
  used e.g.\ in the Baikov representation, we have
\begin{center}
\begin{lstlisting}
  tmpids = CATInvMonIdsFromAnnihilators[$\ldots$];
  deriv  = CATInvMonIdsFromDiffOperators[$\ldots$];
\end{lstlisting}
\end{center}
\item For \emph{duals} of loop integrals, with $\varphi_\alpha$ defined as in section~\ref{sec:duals}, i.e.\ implementing eq.~\eqref{eq:dualmon} and~\eqref{eq:dualdiff}, we need to first specify which denominators are regulated (i.e.\ \emph{can} appear with positive exponent) using
\begin{center}
\begin{lstlisting}
  CATDualRegulated[$F$] = {$\sigma_1$,$\ldots$,$\sigma_n$};
\end{lstlisting}
\end{center}
where $\sigma_j=1$ ($\sigma_j=0$) if the $j$-th denominator is regulated (not regulated).  Then, the identities are found with
\begin{center}
\begin{lstlisting}
  tmpids = CATDualInvMonIdsFromAnnihilators[$\ldots$];
  deriv  = CATDualInvMonIdsFromDiffOperators[$\ldots$];
\end{lstlisting}
\end{center}
\item For Mellin moments with $\varphi_\alpha$ defined as in eq.~\eqref{eq:mellin} we have
\begin{center}
\begin{lstlisting}
  tmpids = CATMellinIdsFromAnnihilators[$\ldots$];
  deriv  = CATMellinIdsFromDiffOperators[$\ldots$];
\end{lstlisting}
\end{center}
\end{itemize}
The Mellin integrals defined in eq.~\eqref{eq:mellin}, while closely related to loop integrals, generally differ from them by an $\alpha$-dependent prefactor. It is generally convenient to convert the template identities and rewrite them directly in terms of loop integrals.  We achieve this in \textsc{CALICO} by using, instead of the code snippet of the previous block,
\begin{center}
\begin{lstlisting}
  tmpids = CATSchwingerIdsFromAnnihilators[$F$,ann,$\z$];
  deriv  = CATSchwingerIdsFromDiffOperators[$F$,diffop,$\z$];
\end{lstlisting}
\end{center}
for the Schwinger representation and
\begin{center}
\begin{lstlisting}
  tmpids = CATLPIdsFromAnnihilators[$F$,ann,$\z$,$\ell$,$d$];
  deriv  = CATLPIdsFromDiffOperators[$F$,diffop,$\z$,$\ell$,$d$];
\end{lstlisting}
\end{center}
for the Lee-Pomeransky representation.  Note that, for the latter, the two functions also take, as input, the number of loops $\ell$ and space-time dimensions $d$, since the prefactors to be adjusted depend on them.

Once template identities are available, we typically want to apply them to a list of seeds to generate systems of identities to be solved, as explained below eq.~\eqref{eq:Atmpeq}.  Generating seeds and the corresponding system is up to the user.  A typical strategy involves generating, for all sectors $S_\beta$ (with $\beta_j=0,1$, defined as in section~\ref{sec:loop}) all seeds up to a certain numerator degree $s$ (or rank) and a certain number of dots $u$.  \textsc{CALICO} can help to obtain this list via
\begin{center}
\begin{lstlisting}
  CATGenerateSeeds[$\beta$,{$u_{\textrm{min}}$,$u_{\textrm{max}}$},{$s_{\textrm{min}}$,$s_{\textrm{max}}$}]
\end{lstlisting}
\end{center}
which generates all seeds $\alpha$ such that $I_\alpha\in S_\beta$ and $u_{\textrm{min}}\leq u \leq u_{\textrm{max}}$, $s_{\textrm{min}}\leq s \leq s_{\textrm{max}}$.

Once the system of identities is generated from the template identities, we need to choose an ordering, or a weight (see section~\ref{sec:ann}), in order to solve it. It is generally useful to list all integrals appearing in an expression (for instance, a system of equations), sorted by weight. This is achieved with
\begin{center}
\begin{lstlisting}
  CATIntCases[expression,weight]
\end{lstlisting}
\end{center}
which returns a list of integrals, identified by the symbolic function \texttt{CATInt}, sorted by the specified weight as if using
\begin{center}
\begin{lstlisting}
  SortBy[$\textrm{``list of integrals in the expression''}$,weight]
\end{lstlisting}
\end{center}
with \texttt{weight} being a suitable function.  \textsc{CALICO} includes several implementations for the function \texttt{weight} tailored to loop integrals. Here we list them with the most important criteria they use:
\begin{itemize}
\item \texttt{CATIntMellinDefaultWeight} eliminates numerators, i.e.\ negative exponents, of loop integrals, and it is more conveniently used with the representations of loop integrals as twisted Mellin transforms, described in section~\ref{sec:twisted};
\item \texttt{CATIntInvMonDefaultWeight} is conveniently used with Baikov representations (see section~\ref{sec:baikov}) and prefers integrals with lower $u$, or lower $t$ as a tie-breaker;
\item \texttt{CATIntDualInvMonDefaultWeight} is conveniently used with duals of loop integrals in Baikov representations (see section~\ref{sec:duals}) and still prefers integrals with lower $u$, but prefers higher values of $t$ as a tie-breaker.
\end{itemize}
These functions sort higher-weight integrals on the left, as required when using the linear solver of \textsc{FiniteFlow} --- although any other solver or definition of weight can be used.

As already stated, the integrals in a family may also obey additional relations, such as symmetry relations, which cannot be found by the methods described here and thus have to be found and implemented by the user.

We refer to the builtin examples for a description of how to combine all these ingredients in several applications.

Finally, \textsc{CALICO} also exposes functions for solving syzygy equations, since this can be useful for a much broader set of problems, both within and outside physics.  Generators of syzygy solutions for eq.~\eqref{eq:syz} up to a maximum degree $d_{\textrm{max}}$ can be obtained using
\begin{center}
\begin{lstlisting}
  g = CATSyz[{$f_1(\z)$,$\ldots$,$f_k(\z)$},{$z_1,\ldots,z_n$},$d_{\textrm{max}}$]
\end{lstlisting}
\end{center}
which can also take most of the options of \texttt{CATAnnihilator}.  The returned value \texttt{g} is a list which contains, at position $d+1$, all the generators of degree $d$ that have been found.  Each generator is in turn a list of polynomials $\g(\z)$ that satisfy the syzygy equation.

Similarly, a solution $\g(\z)$ for the polynomial decomposition in eq.~\eqref{eq:polydec} can be obtained using
\begin{center}
\begin{lstlisting}
  g = CATPolyDec[{$f_1(\z)$,$\ldots$,$f_k(\z)$},{$h_1(\z)$,$\ldots$,$h_m(\z)$},{$z_1,\ldots,z_n$},$d_{\textrm{max}}$]
\end{lstlisting}
\end{center}
which returns a list of length $m$, whose $i$-th entry is a solution of eq.~\eqref{eq:polydec} for $h(\z)=h_i(\z)$, or \texttt{CATImpossible} if no such solution was found.

\subsection*{Advanced usage}

Users who wish to implement custom identities that are not already supported by the package may need to understand the form of the output of the function \texttt{CATAnnihilator}. The format is
\begin{center}
\begin{lstlisting}
  {
    {$A_{10}$,$A_{11}$,$\ldots$,$A_{1 d_{\textrm{max}}}$},
    {$A_{20}$,$A_{21}$,$\ldots$,$A_{1 d_{\textrm{max}}}$},
    $\ldots$,
    {$A_{o_{\textrm{max}} 0}$,$A_{o_{\textrm{max}} 1}$,$\ldots$,$A_{o_{\textrm{max}} d_{\textrm{max}}}$}
  }
\end{lstlisting}
\end{center}
where $A_{od}$ is the \emph{list} of generators of order $o$ and degree $d$ which have been found.  Each element $A_{od}[[k]]$ of $A_{od}$ is itself a list containing the coefficients $c_{j_1 j_2\cdots}(\z)$ which define the annihilator as in eq.~\eqref{eq:Ao}.  More precisely, the $i$-th element of the list $A_{od}[[k]]$ is the polynomial which multiplies the differential operator
\begin{equation}
  \prod_{j=1}^n(\partial_j)^{a_{ij}}\label{eq:diffopexplicit}
\end{equation}
where $\{a_{i 1},\ldots,a_{i n}\}$ is the $i$-th element of the list returned by
\begin{center}
\begin{lstlisting}
  Join@@Table[CATExponentList[$o$,$n$],{$o$,0,$o_{\textrm{max}}$}]
\end{lstlisting}
\end{center}

Similarly, more advanced use cases may require a better understanding of the output of \texttt{CATDiffOperator}.  As already stated, this is a list of length $m$, where $m$ is the number of free parameters which we need differential operators for.  Each entry of this list corresponds to the solution for the $i$-th parameter and contains either the symbol \texttt{CATImpossible}, to signal no solution was found, or a rule of the form
\begin{center}
\begin{lstlisting}
  $o$ -> {$c_0^{(x)}(\z)$,$c_1^{(x)}(\z)$,$\ldots$}
\end{lstlisting}
\end{center}
where $o$ is the order of the operator that has been found and the list on the right contains the coefficients $c^{(x)}_{j_1 j_2\cdots}(\z)$ that define the operator as in eq.~\eqref{eq:Oo}.  In particular, the $i$-th element of such a list is the polynomial that multiplies the differential operator defined as in eq.~\eqref{eq:diffopexplicit}.

As already stated, the twist $u(\z)$ taken as input by the functions above can be passed just by using its analytic expression.  It is, however, sometimes convenient to use symbolic polynomials, inside the twist, in a first stage of the analytic preparation performed by such functions, then substitute their analytic expressions only at a later stage.  For this purpose, one can specify the twist in the following form
\begin{center}
\begin{lstlisting}
  CATTwist[$u(\z,\x)$,{$B_1$->$\ldots$,$B_2$->$\ldots$,$\ldots$}]
\end{lstlisting}
\end{center}
where $\z$ are the integration variables and $\x$ are the list of free parameters, while $u(\z,\x)$ is an expression for the twist where all polynomials that define it are replaced by \emph{symbolic functions} of the form $B_i(\z,\x)$.  The second argument is a replacement rule or a list of them that is used to replace the polynomial functions $B_i$ with their explicit expression.  Note that this replacement will be done on expressions that depend on both the polynomials $B_i(\z,\x)$ and their (first or higher-order) derivatives with respect to $z_j$ --- and also with respect to $x_j$ when computing operators $\hat O_{x_j}$.  Hence, the replacement should involve the \emph{heads} $B_i$ rather than the functions evaluated at specific arguments $(\z,\x)$.  All twists computed by \textsc{CALICO}'s functions are in this form, hence we suggest examining any of those for explicit examples.

\section{Conclusions and outlook}
\label{sec:conclusions}

In this paper we reviewed and expanded on the method of parametric annihilators for finding linear relations among functions having a suitable integral representation.  We illustrated this approach in a way that applies to a broad class of integral families. These include loop integrals in several parametric representations.
We applied it to examples involving hypergeometric functions, loop integrals in the Lee-Pomeransky or Baikov representation (for which this or similar approaches had already been formulated~\cite{Larsen:2015ped,Bitoun:2017nre}), as well as to the loop-by-loop Baikov and Schwinger representations which, until now, had not been systematically used for the purpose of integral reduction. Building on similar principles, we illustrated a method to derive differential equations for the independent master integrals. We also showed that our formulation applies to duals of loop integrals, which play a crucial role in intersection theory.

We described algorithms for computing annihilators and the differential operators needed for differential equations which exploit modern sparse linear solvers and finite-field techniques. We released an implementation of it, in the public \textsc{Mathematica} package \textsc{CALICO}. At the time of writing, the \textsc{CALICO} package can be useful for theoretical studies involving various representations of loop integrals and special functions.

In the future, on top of implementing additional optimizations, we plan to release an interface between it and the \textsc{FFIntRed} package (also to be published) for the systematic generation of identities needed for the reduction of loop integrals contributing to a process. \textsc{CALICO} numerically computes and analytically reconstructs the analytic dependence of differential operators on external parameters using nodes of \textsc{FiniteFlow}'s computational graphs. In the future, we plan to add an interface which enables using these nodes in a custom graph to generate identities. This would sidestep the need to reconstruct complex template identities for multiscale processes, while still benefitting from this method for computing linear relations.

We believe the techniques and applications illustrated in this paper, as we as the \textsc{CALICO} package we released, could become highly beneficial to numerous future applications, including studies that rely on the strengths of several representations of loop integrals.

\section*{Acknowledgements}
We thank Vsevolod Chestnov, Thomas Gehrmann, Pierpaolo Mastrolia, Kay Sch\"{o}nwald, Simone Zoia and Lorenzo Tancredi for feedback on our work and on the draft. The work of GF received funding from the European Research Council (ERC) under the European Union’s Horizon 2020 research and innovation programme grant agreement 101019620 (ERC Advanced Grant TOPUP) and from the UZH Candoc scheme (Candoc grant Nifty LooPS).  The work of TP received funding from the European Research Council (ERC) under the European Union’s Horizon Europe research and innovation programme grant agreement 101040760 (ERC Starting Grant FFHiggsTop).

\appendix

\section{Representations of loop integrals}
\label{sec:feynLPpar}

In this appendix, we review the definition of the polynomials appearing in the twist $u(\z)$ of the representations of loop integrals used in the paper.  In the following, we consider an $\ell-$loop Feynman integral with $e+1$ external legs --- out of which only $e$ are independent because of momentum conservation. Loop integrals are defined as in eq.~\eqref{eq:loopint}, with $n$ generalized denominators with the form in equations~\eqref{eq:loopden} or~\eqref{eq:loopdenbl}. In the following $\z=(z_1,\ldots, z_n)$ is the list of integration variables that these polynomials depend on, in the parametric representations we consider.

\subsection*{Baikov polynomial}
The Baikov representation results from a change of variables from the $d$-dimensional loop momenta to the generalized denominators. By inverting equations~\eqref{eq:loopden} and~\eqref{eq:loopdenbl}, we rewrite every scalar product of the form $k_i\cdot k_j$ or $k_i\cdot p_j$ as a linear combination of generalized denominators
\begin{align}
  k_i\cdot k_j ={}& d_{ij0} + \sum_{m=1}^n d_{ijm}\, D_m \nn
  k_i\cdot p_j ={}& e_{ij0} + \sum_{m=1}^n e_{ijm}\, D_m , \label{eq:sptoden}
\end{align}
where $d_{ijm}$ and $e_{ijm}$ are functions of external invariants. This inversion requires a full set of $n$ generalized denominators, with $n$ given in eq.~\eqref{eq:fulln}. The Baikov polynomial is a Gram determinant. We recall that, given a list of vectors $\{v_j\}_{j=1}^m$, their Gram determinant is defined as
\begin{equation}
  \text{Gram}(v_1,\ldots,v_m) = \text{det} V, \qquad \textrm{with } V_{ij} \equiv 2\, v_i\cdot v_j.
\end{equation}
The Baikov polynomial $B(\z)$ is given by the Gram determinant of the loop momenta $k_j$ and external momenta $p_j$, after rewriting each scalar product involving loop momenta as a linear combination of generalized denominators (as in eq.~\eqref{eq:sptoden}) and replacing each denominator $D_j$ with the variable $z_j$.  More explicitly,
\begin{equation}
B(\z) = \text{Gram}(k_1,\ldots k_\ell,p_1,\ldots,p_e) \Big|_{ \substack{ k_i\cdot k_j\, \to\, d_{ij0} + \sum_{m=1}^n d_{ijm}\, z_m \nn
  k_i\cdot p_j\, \to\, e_{ij0} + \sum_{m=1}^n e_{ijm}\, z_m}}.
\end{equation}
This polynomial enters the Baikov representation in equations~\eqref{eq:JtoIbaikov} and~\eqref{eq:baikov}.

We also add that the proportionality factor $K$ in eq.~\eqref{eq:JtoIbaikov} is proportional to another Gram determinant, namely
\begin{equation}
  K = C(d)\, \text{Gram}(p_1,\ldots,p_e)^{\frac{-d+e+1}{2}}.
\end{equation}
In the standard Baikov representation, this prefactor is irrelevant in linear integral identities, since it is the same for all integrals within a family. In the loop-by-loop Baikov representation~\cite{Frellesvig:2017aai} --- which consists in applying the Baikov parametrization one loop at the time --- the external legs of a subloop generally depend on the other loop momenta. The Gram determinant in the last relation thus contributes to the twist in this case, except for the one corresponding to the last loop integration that is rewritten in this parametric form. Hence, in the loop-by-loop Baikov representation, we obtain a twist $u(\z)$ with the form in eq.~\eqref{eq:u} with $2\ell-1$ polynomials $B_j(\z)$ and exponents $\gamma_j$, where the latter are linear functions of $d$.  The external prefactor, which we wrote as $B_0^{\gamma_0}$ in the example in eq.~\eqref{eq:lbluexample}, must however be included when finding DEs, e.g.\ via differential operators $\hat O_x$ as described in section~\ref{sec:de}, since it depends on the external invariants $x$.

\subsection*{Symanzik polynomials}
The first and second Symanzik polynomials, respectively denoted as $\U(\z)$ and $\F(\z)$, appear in the twists of both Lee-Pomeranski~\eqref{eq:LP} and Schwinger~\eqref{eq:schwinger} representations. They also appear in the well-known Feynman parametrization. We briefly recall how they can be computed algebraically. Given the structure of the generalized denominators $D_j$, the sum of the denominators $D_j$ weighted by the variables $z_j$ has the following dependence on the loop momenta,
\begin{equation}
  \sum_{j=1}^n z_j\, D_j = \sum_{i,j=1}^n A_{ij}\, (k_i \cdot k_j) + 2 \sum_{j=1}^{\ell} (B_j \cdot k_j) + C,
\end{equation}
which defines the symmetric matrix $A_{ij}$, the Lorentz vectors $B_j^{\mu}$ and the scalar $C$, all of which have a linear dependence on $\z$.  The vector $B_j^\mu$ and the scalar $C$ also depend on the external kinematics.
The polynomials $\U(\z)$ and $\F(\z)$ are thus obtained as
\begin{align}
  \U(\z) &= \text{det} A,\\
  \F(\z) &= \U(\z) \left(\sum_{i,j=1}^n A_{ij}^{-1} (B_i \cdot B_j) - C\right),
\end{align}
with $A^{-1}$ being the inverse of the matrix $A$.  Note that $\text{det} A \times A^{-1}$ is a polynomial function of $A_{ij}$, hence $\F(\z)$ is also polynomial in $\z$.

\bibliographystyle{JHEP}
\bibliography{biblio}

\end{document}